\documentclass[]{aa}  

\usepackage{graphicx}
\usepackage{txfonts}
\usepackage{xcolor}
\usepackage{amsmath}
\usepackage{amsfonts}
\usepackage{natbib}
\usepackage{longtable}
\usepackage{booktabs}
\usepackage{multirow}
\usepackage[amsmath,thmmarks]{ntheorem}
\usepackage[normalem]{ulem}
\theoremseparator{.}
\theorembodyfont{\normalfont} 
\newtheorem{assumption}{Assumption}{\bfseries}{\itshape}

\usepackage[normalem]{ulem}
\usepackage{hyperref}
\hypersetup{
    colorlinks=true,
    linkcolor=blue,
    filecolor=magenta,      
    urlcolor=cyan,
    citecolor=teal,
    pdftitle={},
    }
\begin{document}

   \title{Going Bayesian on the ages of nearby young stellar systems}
   \subtitle{I. The expansion rate method}

   \author{J. Olivares\inst{1} 
   \and A. Berihuete\inst{2}
   \and H. Bouy\inst{3}
   }
   
   \institute{
   		Departamento de Inteligencia Artificial, Universidad Nacional de Educación a Distancia (UNED), c/Juan del Rosal 16, E-28040, Madrid, Spain. 
   		\email{jolivares@dia.uned.es}
		\and
		Depto. Estadística e Investigación Operativa, Universidad de Cádiz, Avda. República Saharaui s/n, 11510 Puerto Real, Cádiz, Spain.
		\and
		Laboratoire d'astrophysique de Bordeaux, Univ. Bordeaux,CNRS, B18N, allée Geoffroy Saint-Hilaire, 33615 Pessac, France.
        }

   \date{}

 
  \abstract
   {Determining the ages of young stellar systems is fundamental to test and validate current star-formation theories.}
   {We aim at developing a Bayesian version of the expansion rate method that incorporates the a priori knowledge on the stellar system's age and solves some of the caveats of the traditional frequentist approach.} 
   {We upgrade an existing Bayesian hierarchical model with additional parameter hierarchies that include, amongst others, the system's age. For this later, we propose prior distributions inspired by literature works.}
   {We validate our method on a set of extensive simulations mimicking the properties of real stellar systems. In stellar associations between 10 and 40 Myr and up to 150 pc the errors are <10\%. In star forming regions up to 400 pc, the error can be as large as 80\% at 3 Myr but it rapidly decreases with increasing age. }
   {The Bayesian expansion rate methodology that we present here offers several advantages over the traditional frequentist version. In particular, the Bayesian age estimator is more robust and credible than the commonly used the frequentist ones. This new Bayesian expansion rate method is made publicly available as a module of the free and open-source code \textit{Kalkayotl}.}


   \maketitle
%
\section{Introduction}
\label{introduction}

Accurate and precise age determinations of young stellar systems are fundamental elements for the validation of star-formation theories. To this end, the nearby and young stellar systems (NYSS) are fundamental benchmarks because they offer the best quality data and are more likely to preserve the initial conditions of the star-formation processes. However, inferring the age of these systems is a difficult task due to the often inconsistent results obtained from different dating techniques \citep[see, for example,][]{2022A&A...664A..70G,1999ApJ...522L..53B,1995AJ....109..298S}.

Among the various dating techniques that could be use in NYSS, dynamical ages, such as the expansion rate method \citep[e.g.][]{2024MNRAS.533..705W,2014MNRAS.445.2169M,1997MNRAS.285..479B} and traceback method \citep[e.g.][]{2023MNRAS.520.6245G,2022A&A...667A.163M} are fundamental because they provide independent age estimates from those given by other methods, like isochrone-fitting or Lithium depletion boundary \citep[see, for example,][and references therein]{2023A&A...678A..71R,2023MNRAS.523..802J,2022A&A...664A..70G,2006ApJ...645.1436V}. 

The dynamical dating techniques make the underlying assumption that young stellar systems were more spatially compact at their birth time and that their current size can be mapped to their birth size using the current positions and velocities of their members. The systems' age is thus the time needed for its members to pass from their original configuration to the current one. In the traceback method, the mapping is done by tracing back in time the system's members current positions based on their velocities and assuming a Galactic potential. On the other hand, in the expansion rate method, the mapping is done assuming that the current positions and velocities of the system's members are the consequence of a free expansion at a constant rate. This rate measures the average change in the members' velocity with respect to the distance to the system's centre \citep[see, for example, the section "The concept of linear expansion" in][]{1964ARA&A...2..213B}. Thus, the expansion rate is an spatial rate rather than a temporal one.

In the expansion rate dating method, the expansion rate, $\kappa$, is inverted to obtain an estimate of the stellar system's age, $\tau$, through the following equation, 
\begin{equation}
\label{equation:expansion_to_age}
\tau = (1.02271\kappa)^{-1},
\end{equation} 
with $\tau$ and $\kappa$ in units of Myr and $\rm{km\,s^{-1}\,pc^{-1}}$, respectively.

We notice that this method makes the underlying assumption that the stellar system is expanding, that this expansion originated at the system's birth, and that it has remained constant since then (see Assumptions \ref{assumption:expansion}, \ref{assumption:expansion_at_birth}, and \ref{assumption:constant_expansion}).

In this work we revisit the expansion rate method and provide the community with a Bayesian implementation of it in the form of a module to the free and open-source code \textit{Kalkayotl} \citep{2025A&A...693A..12O}. The statistical problem of inverting the probability distribution of the expansion rate to obtain a distribution of the age share similarities with that of inverting parallaxes to obtain distances \citep{2015PASP..127..994B}, and thus, it strongly benefits from a Bayesian approach, particularly in the domain of low signal to noise, as is the case of the expansion rates of NYSS.

The rest of this work is organised as follows. In Sect. \ref{data}, we present the synthetic datasets that we will use to validate our new Bayesian dating method. Section \ref{methods}, describes the traditional frequentist expansion rate dating methods and their caveats and introduces our new Bayesian approach with its various priors and hyper-parameters. Then, in Sect. \ref{sensitivity}, we do a series of sensitivity analyses to select the optimal configurations for the prior distributions and hyper-parameters of our model that maximise its performance. Afterwards, in Sect. \ref{validation}, we validate the optimal configuration of our method with synthetic datasets of NYSS. Finally, in Sects. \ref{discussion} and \ref{conclusions}, respectively, we discuss the advantages and limitations of our new Bayesian method and present our conclusions and general recommendations for its use. In addition, Appendix \ref{appendix:assumptions} provides a comprehensive list of the assumptions we undertake in the construction of our new dating method together with critics to each of them.

\section{Data}
\label{data}

In this Section, we present the data that will be used to validate our Bayesian expansion rate method. We start by establishing the parameter space of the method, this is, its applicability domain, together with a grid of sensible values for it. Afterwards, we describe the characteristics of the synthetic data generation mechanism. Finally, we describe the properties of the synthetic systems that will remain fixed (i.e. those not in the grid of parameter space values), which will be based on archetypical ones.

\subsection{Applicability domain}
\label{data:applicability_domian}

The applicability domain of the expansion rate method has been poorly characterised in the literature. \citet{1997MNRAS.285..479B} did the the first characterization of the caveats of a version of the expansion rate dating method in which the expansion rate of the stellar system is determined from the rate of change of the proper motions of its members with respect to their sky positions. Those authors conclude that the kinematic ages determined with this method are limited by the uncertainties in the proper motions and parallax, and that the ages thus determined can be underestimated or overestimated depending on the chosen coordinate direction. Furthermore, they found that their ages eventually converge towards a value $\sim$4 Myr for associations older that this age, thus indicating that the applicability domain of their expansion rate method is age dependent and saturates at 4 Myr. Recently, \citet{2019A&A...628A.123Z} determine kinematic ages in Orion using a restricted version of \citet{2000A&A...356.1119L} model in which the expansion is assumed isotropic and does not include rotations or shear terms. In their Appendix A.2, those authors use simulated data mimicking the \textit{Gaia} properties to show that their model is able to retrieve the expansion parameter in a Hyades like cluster but underestimate it in an Orion like one, thus indicating that the applicability domain of the expansion rate method is distance dependent. These two previous works indicate that the applicability domain of the expansion rate method depends on the age and distance of the stellar system.

In this work, we will infer expansion ages through a forward modelling method in which the age, expansion rates, positions and velocity dispersions are inferred from an extension to the linear velocity field model of \citet[][see Sect. \ref{methods}]{2000A&A...356.1119L}. This Bayesian hierarchical model with linear velocity field is implemented as part of the \textit{Kalkayotl} code \citep{2025A&A...693A..12O,2020A&A...644A...7O}, which as been thoroughly validated through a set of synthetic data simulations mimicking the properties of \textit{Gaia} DR3 data. The results of those simulations, particularly those pertaining to the detectability of the linear velocity field entries \citep[see Sect. 3.1.2 of][]{2025A&A...693A..12O}, show the the signal-to-noise ration of the expansion rate components not only depends on the age (i.e. the strength of the expansion signal) and distance of the system, but also in its number of stars. Moreover, the three values of the $C$ constant shown in Figure 3 of the aforementioned authors can be converted to expansion ages using Eq. \ref{equation:expansion_to_age}, resulting in ages of $\sim$100 Myr ($C\!=\!10\,\rm{m\,s^{-1}\,pc^{-1}}$), $\sim$20 Myr ($C\!=\!50\,\rm{m\,s^{-1}\,pc^{-1}}$), and $\sim$10 Myr ($C\!=\!100\,\rm{m\,s^{-1}\,pc^{-1}}$), which correspond to the typical ages of young open clusters, stellar associations, and star-forming regions, respectively. Therefore, from this figure we can conclude that even with a very relaxed threshold of a S/N of 1, the expansion rate method would not work in open clusters.

Based on the previous analysis, we conclude that a sensible parameters space to probe the applicability domain of the expansion rate method could be split in two regimes. On one hand, stellar associations with ages between 10 Myr and 50 Myr and distances up to $\sim$200 pc. On the other hand, star-forming regions with ages <10 Myr and distances up to $\sim$500 pc. The typical populations of stellar associations vary between 20 and 80 members \citep[see, for example, Table 1 of ][]{2018ApJ...856...23G} while those of star-forming regions are typically of a few hundreds (see, for example, the cases of $\rho$-Ophiuchi and Taurus in the aforementioned table).

\subsection{Generation of synthetic stellar systems}
\label{data:synthetic_data_generation}

We generate synthetic datasets with the \textit{Amasijo}\footnote{\url{https://github.com/olivares-j/Amasijo}} code \citep[see for example][]{2023A&A...675A..28O,2022A&A...664A..31C}. This free and open-source code takes as input astrometric and photometric parameters together with a random seed generator and the total number of desired system members. These inputs are used to generate the phase-space coordinates and masses of the stellar system's members with a variety of statistical distributions. Here, we assume uniform and multivariate normal distributions for the masses and phase-space coordinates, respectively. A normal distribution suffices to describe the phase-space distribution of the majority of the known stellar associations and subgroups within star-forming regions. We choose to sample masses from a uniform distribution rather than other forms of the initial mass function to directly constrain the fraction of sources with observed radial velocity by means of a lower mass limit that translates into a photometric limit. In dynamical age determinations \citep[see for example][]{2023MNRAS.520.6245G,2020A&A...642A.179M,2019A&A...628A.123Z}, the coverage of radial velocities is more crucial than the completeness of the mass function.

Once the phase-space coordinates and masses were generated, they are transformed to the observable space using \textit{PyGaia}\footnote{\url{https://github.com/agabrown/PyGaia}} and  \textit{isochrones}\footnote{\url{https://github.com/timothydmorton/isochrones}}  \citep{2015ascl.soft03010M}. The \textit{isochrones} code generates the true values of the \textit{Gaia} observed photometry based on the source's mass, distance, metallicity, extinction, and input age. Throughout this work we set the metallicity to the solar value and extinction to zero. These choices reflect the values of the majority of the stellar associations in the solar neighbourhood. In the case of star-forming regions, a higher extinction would be roughly equivalent to a degradation of the astrometric uncertainty, and thus, will be similar to placing the system at a larger distance.

\textit{Amasijo} then uses the \textit{PyGaia} code to obtain the observational uncertainties for the desired \textit{Gaia} data release according to the true astrometric, photometric, and radial velocity values. Finally, the observed values are sampled from univariate normal distributions in which the true values are the mean and the uncertainties the standard deviations. The output of the code consists of a comma-separated-value file with the observed and true values of the generated sources, and a plot with a summary of the synthetic data properties.

Additional parameters of the \textit{Amasijo} code are the fraction of sources with observed radial velocities, the maximum phase-space Mahalanobis distance of the sources, and a magnitude shift to increase the astrometric uncertainties. The synthetic astrometric uncertainties provided by \textit{PyGaia} are much smaller than those observed in real stellar stellar systems, thus, we applied this magnitude shift to increase the astrometric uncertainties and match the real ones. The maximum phase-space Mahalanobis distance is useful to avoid sources that by simple random effects got generated at unplausible large phase-space distances. Throughout this work and in all our simulations, we use a maximum Mahalanobis distance of two, which covers 95\% of the phase-space distribution. We set the uncertainty properties to that of the third \textit{Gaia} data release. The remaining of the parameters were varied with the following grid's values.

Finally, previous to the use of our datasets, we apply the following filtering criteria. First, we filter out sources with parallax smaller than 1.0, this effectively remove distant sources that may have resulted from extreme random number generation. Second, we set a lower limit of 100 $\rm{m\,s^{-1}}$ as the minimum radial velocity uncertainty, and thus, the uncertainty of sources with higher precision were fixed to this minimum value. Third, we apply sigma clipping of sources farther away than three and two standard deviations in parallax and radial velocity, respectively.

\subsection{Parameter's space grid}
\label{data:grid_parameters}

We generate stellar systems mimicking the phase-space parameters of archetypical stellar systems: the $\beta$-Pictoris stellar association and the Taurus star-forming region. We select this benchmark systems due to their proximity and well characterised properties. For $\beta$-Pictoris, we use the phase-space parameters inferred by \citet[][see their Table 4]{2020A&A...642A.179M}, with position dispersion $\sigma_{XYZ}=[16.04,13.18,7.44]$ pc, velocity dispersion $\sigma_{UVW}=[1.49,0.54,0.70]\, \rm{km\,s^{-1}}$, and the nominal (i.e. measured) distance and age values of 51 pc \citep{2020A&A...642A.179M} and $23\pm8$ Myr \citep{2024MNRAS.tmp...30L}, respectively. We choose this latter work because it is the latest one with an age determination that covers the majority of the literature ones. For the Taurus star-forming region, we use the phase-space centroid parameters of the L1495 subgroup inferred by \citet[][see their Tables 2 and 3]{2019A&A...630A.137G}. In this last case, we set the phase-space dispersion to 1 pc in position and 1 $\rm{km\,s^{-1}}$ in velocity, which  correspond to the typical values of the Taurus subgroups and also to those of the NGC1333 and IC348 core subgroups of the Perseus star-forming region \citep{2023A&A...671A...1O}.

\begin{table}[ht!]
\caption{Parameter values for the grid of synthetic simulations}
\label{table:grid_parameters}
\centering
\resizebox{\columnwidth}{!}
{
\begin{tabular}{cccc}
\hline
{} & {} & Stellar associations & Star-forming regions \\
 \hline 
Age & [Myr] & 10, 20, 40 & 3, 5, 7 \\
Distance & [pc] & 50, 100, 150 & 150, 300, 400 \\
Number & {} & 25, 50, 75 & 50, 100 \\
$RV_{fraction}$ & {} & 0.25, 0.50, 1.0 &  0.25, 0.50, 1.0\\
\hline
$\sigma_{XYZ}$ & [pc] & (16.0, 13.2, 7.4) & (1, 1, 1) \\
$\sigma_{UVW}$ & [$\rm{km\,s^{-1}}$] & (1.5, 0.5, 0.7) & (1, 1, 1) \\
\hline
\end{tabular}
}
\end{table}

We replicate the benchmark phase-space templates of $\beta$-Pictoris and Taurus L1495 over the grid of values shown in Table \ref{table:grid_parameters}. We notice that the position and velocity dispersions shown in the last two rows represent the single vector value used to generate the template of either stellar associations star-forming regions, and these values were not varied throughout our analysis except for the experiments shown in Sect. \ref{sensitivity:dispersion}. The ages and distances cover the applicability domain interval  discussed in Sect. \ref{data:applicability_domian}, while the number of sources cover the expected population sizes of the majority of the nearby young stellar associations and star-forming regions  \citep[see, for example, Table 1 of][]{2018ApJ...856...23G}. We use low, medium, and high fraction of radial velocity coverages to test the impact that current and dedicated radial velocity surveys may have on our method. Finally, we avoid drawing conclusions from possible random excursions of the random number generator by generating each grid point five times with different random seeds. We generate the benchmark templates with different ages and distances by shifting the distance vector along its nominal direction and by isotropically fixing the expansion rate components to the requested age using Eq. \ref{equation:expansion_to_age}, respectively.

\section{Methods}
\label{methods}

In this section, we introduce the Bayesian expansion rate method that we propose to infer the ages of NYSS based on their the expansion rate. We start with a review of the current frequentist approach to this dating method. Then, we expose our novel Bayesian alternative approach, together with its associated model and descriptions of its practical implementation and prior distributions. Afterwards, we do sensitivity analyses to identify the optimal combination of the model's prior distributions and hyper-parameters. Finally, we end this section with a comparison of the frequentist and Bayesian approaches to the expansion rate dating method.

\subsection{Frequentist approach}
\label{methods:frequentist}

Traditionally, the frequentist procedure to obtain ages is the following. First, the expansion rates are inferred with maximum likelihood methods \citep[e.g.][]{2023AJ....165..269L,2022AJ....164..151L,2019A&A...628A.123Z,2018MNRAS.476..381W,2014MNRAS.445.2169M,2002A&A...387..117B}, where the fits are performed directly on the observed spaces \citep[i.e. proper motions and sky positions, e.g.][]{2002A&A...387..117B}, in the physical ones \citep[i.e. space positions and velocities, e.g.][]{2023AJ....165..269L,2022AJ....164..151L,2019A&A...628A.123Z,2014MNRAS.445.2169M} or in combinations of them \citep[e.g.][]{2018MNRAS.476..381W}. These fits are independently performed in one or several orthogonal directions (e.g. radial velocity versus distance or $U$ vs $X$, $V$ vs $Y$, and $W$ vs $Z$) with the correlations between these directions usually disregarded. It can also be assumed that the expansion is uniformly isotropic, as done by \citet{2019A&A...628A.123Z}. If the procedure results in more than one expansion rate component, then, a variety of criteria are used to combine them. These criteria go from discarding components \citep[typically in the $Z$ direction, e.g.][]{2014MNRAS.445.2169M} to mixing them assuming that they are independent and identically distributed (iid). This latter assumption is at the core of the weighted mean procedure, which is typically used to combine the expansion rates of orthogonal directions \citep[e.g.][]{2023AJ....165..269L,2014MNRAS.445.2169M}. Finally, once a single estimate of the expansion rate is derived, the system's age is obtained through Eq. \ref{equation:expansion_to_age}.

The previous traditional approach, to which we refer hereafter as the classical frequentist method (represented with the symbol $\hat{\vec{\kappa}}$), suffers from the following caveats. First, it underestimates uncertainties. The fact that the expansion rate vector, $\vec{\kappa}$, can be decomposed in its orthogonal components $\vec{\kappa}\!=\kappa_X\cdot\vec{e}_X+\kappa_Y\cdot\vec{e}_Y + \kappa_Z\cdot\vec{e}_Z$, does not necessarily mean that these components are statistically uncorrelated. As it is widely known, neglecting correlations result in underestimated uncertainties \citep[see Sect. 14.7 of][]{2014bda..book.....G}. Second, the inversion of a noisy estimate of the expansion rate, $\hat{\kappa}$, may result in a biased age estimate in a similar way in which the inversion of a noisy estimate of the parallax results in a biased distance estimate \citep{2015PASP..127..994B}. 

The last caveat of the classical method can be partially mitigated by combining and inverting only those components of the expansion rate with sufficient signal-to-noise ratio, for example S/N>3. Hereafter, we will refer to this latter approach as the robust frequentist method. We notice that the robust method still suffers from underestimated uncertainties due to neglect of correlations.

\subsection{Bayesian approach}
\label{methods:bayesian}

In our perspective, the problem of inverting the expansion rate to derive an expansion age has a strong statistical resemblance to the problem of inferring an object distance by inverting its observed parallax. As demonstrated by the works of \citet{2015PASP..127..994B,2023AJ....166..269B} and \citet{2021AJ....161..147B,2018AJ....156...58B} the Bayesian approach offers strong advantages to the distance inference problem. Therefore, in the following, we will pose the expansion age inference problem under a Bayesian framework in which we will address the caveats of the traditional approach: neglected correlations, underestimated uncertainties, and biased ages.

We use the Bayesian hierarchical modelling formalism \citep{2014bda..book.....G} to propagate data correlations and uncertainties towards the expansion rate components (i.e. $\kappa_X$, $\kappa_Y$ and $\kappa_Z$). Then, these latter are pooled together and inverted to obtain a single age estimate. 

\subsubsection{Bayesian expansion age model}
\label{methods:age_model}
In the statistical generative model framework, the age parameter, $\tau$, is inferred as a random variable, which is sampled from its prior (more below), and transformed into the expansion rate, $\kappa_{\mu}$, using the relation $\kappa_{\mu} \, \rm{[km\,s^{-1}\,pc^{-1}]}=(1.02271\, \tau \, \rm{[Myr]})^{-1}$. Then, this $\kappa_\mu$ is used as the location parameter of a statistical distribution (more below) from which the three $\vec{\kappa}$ components (i.e. $\kappa_X$, $\kappa_Y$, and $\kappa_Z$) are sampled. These components are then inserted into the \citet{2000A&A...356.1119L} linear velocity model,

\begin{equation}
\label{equation:linear_model}
\vec{v}_i = \left(\begin{array}{ccc}
\kappa_X & \Omega_{00} & \Omega_{01} \\ 
\Omega_{10} & \kappa_Y & \Omega_{02} \\ 
\Omega_{11} & \Omega_{12} & \kappa_Z
\end{array} \right)\cdot(\vec{x}_i-\vec{x}_0)+ \mathcal{N}(\vec{v}_0,\Sigma_{\vec{v}}).
\end{equation}
in which $\vec{x}_i$ and $\vec{v}_i$ are the position and velocity vectors of member $i$ in the reference frame of the stellar system, $\vec{x_0}$ and $\vec{v_0}$ are the position and velocity vectors of this later, $\Sigma_v$ is the velocity dispersion covariance matrix of the assumed multivariate-normal distribution $\mathcal{N}$, and $\Omega$ is the set of velocity tensor entries related to the system's rotation and shear. The resulting extended linear velocity field model is implemented in the 6D version of \textit{Kalkayotl} \citep{2025A&A...693A..12O} where the posterior distribution of all its parameters, including the system's expansion age $\tau$, $\kappa_\mu$, $\kappa_\sigma$, and other additional parameters, is sampled using the NUTS algorithm \citep{2011arXiv1111.4246H} within the PyMC environment \citep{pymc2023}. The diagnostics and summaries of the specific age model parameters are seamlessly incorporated in the standard output of the \textit{Kalkayotl} code.

The hierarchical approach described above solves the two caveats of the traditional frequentist approach in the following ways. First, by drawing the three expansion rate components from the same parent distribution and sampling the parameters of this latter with the NUTS method their correlations are directly propagated towards the parameters of the upper hierarchies. Second, the pooling of the three components into a single parameter $\kappa_\mu$, increases the statistical significance of the latter and diminishes the bias that may be incurred when inverting it. Finally, when the parameters of a Bayesian hierarchical model are sampled with the NUTS method, the data uncertainties are fully propagated towards the parameters in upper hierarchies, particularly towards the age parameter. This full propagation avoids the typical problem of uncertainty leak that occurs when successive trimming is applied to the tails of the parameter's distributions when they are propagated through the chain of steps involved in the expansion age computation. For example, it is typically assumed that the uncertainties in the phase-space coordinates are Gaussian, that the uncertainties in the expansion rate components are Gaussian, and that the inversion of a Gaussian distribution results in a Gaussian one, none of which is necessarily true. 

\subsubsection{Prior distributions}
\label{methods:prior}

We now describe the prior distributions and parametrization used for the additional age model parameters as well as additional marginal prior distributions for the position and velocity dispersions: $\vec{\sigma}_{XYZ}$ and $\vec{\sigma}_{UVW}$ (see Table \ref{table:prior}). The rest of the linear velocity model parameters and their prior distributions are left untouched. 

\begin{table*}[ht!]
\caption{Prior distribution for the age model parameters.}
\label{table:prior}
\centering
\resizebox{\textwidth}{!}
{
\begin{tabular}{ccccc}
\toprule
Parameter & Parametrization & Distribution & Units & Hyper-parameters \\
\midrule
\multirow[c]{3}{*}{$\tau$} & Central & Generalised Gamma & Myr & location=$\mu_\tau - \sigma_\tau$, scale=$\sigma_\tau$, $d\!=\!p\!+\!1$, $p\!=\!1.19$ \\
& Central & Generalised Gamma & Myr & location=$\mu_\tau - \sigma_\tau \cdot f_\star$, scale=$\sigma_\tau \cdot f_\star$, $d\!=\!p\!+\!1$, $p\!=\!10$ \\
 & Central & Truncated Normal  & Myr & location=$\mu_\tau$, scale=$\sigma_\tau$ \\
$\kappa_\sigma$ & Central & Exponential  & $\rm{km\,s^{-1}\,pc^{-1}}$ & scale=0.01 \\
$\kappa_\nu$ & Central & Gamma  & $\rm{km\,s^{-1}\,pc^{-1}}$ & alpha=2, scale=1 \\
\multirow[c]{2}{*}{$\vec{\kappa}$} & (Non-) Central & Normal & $\rm{km\,s^{-1}\,pc^{-1}}$ & location=$\kappa_\mu$,scale=$\kappa_\sigma$ \\
& (Non-) Central & StudentT & $\rm{km\,s^{-1}\,pc^{-1}}$ & location=$\kappa_\mu$, scale=$\kappa_\sigma$, dof=$\kappa_\nu$ \\
\multirow[c]{3}{*}{$\vec{\sigma}_{XYZ}$} & Central & Truncated Normal & pc & location=20, scale=5 \\
& Central & Gamma & pc & $\alpha$=2.0, $\beta$=1/20 \\
& Central & Exponential & pc & scale=20 \\
\multirow[c]{3}{*}{$\vec{\sigma}_{UVW}$} & Central & Truncated Normal & $\rm{km\,s^{-1}}$  & location=0.5, scale=1.0 \\
& Central & Gamma & $\rm{km\,s^{-1}}$ & $\alpha$=2.0, $\beta$=1/0.5 \\
& Central & Exponential & $\rm{km\,s^{-1}}$ & scale=0.5 \\
\bottomrule
\end{tabular}
}
\end{table*}

The age estimates of NYSS are generally reported in the literature either as intervals or as Gaussian distributions, with a mean and a standard deviation. Therefore, our prior distribution for the age parameter, $\tau$, must have the following two minimum requirements, i) positive support, and ii) free and independent mode and variance parameters. Although there is a large variety of family distributions with positive support (e.g. Gamma, Chi-squared, Exponential, and all the log-type families) their variance and mode are generally linked, thus rendering their independent specification a task that cannot always be solved. For example, in the Chi-square distribution with $k$ degrees of freedom, the mode and the variance are $max(k-2,0)$ and $2k$, respectively, which cannot be set independently. Thus, from the set of family distributions that fulfil the above criteria, we choose to work with the Truncated Normal (on the positive reals) and the Generalised Gamma \citep{Stacy1962}, see Table \ref{table:prior}. 

\begin{figure}[h!]
\centering
\includegraphics[width=\columnwidth]{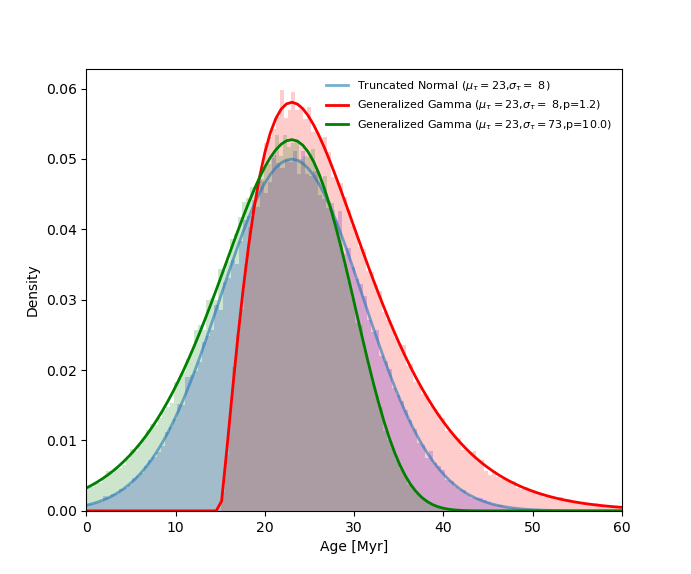}
\caption{Comparison of our age prior distributions. The hyper-parameters ($\mu_\tau$ and $\sigma_\tau$) were fixed to values that result in a distribution with mode and variance corresponding to those for the $\beta$-Pictoris stellar association, $23\pm8$ Myr \citep{2024MNRAS.tmp...30L}.}
\label{figure:prior}
\end{figure}

The Truncated Normal (TN) distribution offers the most direct way to embed as a prior the a priori literature knowledge on the age of a stellar association by directly passing as hyper-parameters the mode, $\mu_\tau$, and standard deviation, $\sigma_\tau$, collected from the literature. On the other hand, the Generalised Gamma distribution has enough flexibility to accommodate varying values of the mode, variance, and skewness. Fixing its parameters to the set of values shown in Table \ref{table:prior} renders distributions with the same mode and variance as a Gaussian distribution, which enable us to directly convey the a priori knowledge from the literature with the mode, $\mu_\tau$, and standard deviation, $\sigma_\tau$\footnote{For the Generalised gamma with negative skewness to have the required standard deviation, the input scale must be multiplied by the factor $f_\star=(\Gamma((d+2)/p))/\Gamma(d/p)-(\Gamma((d+1)/p)/\Gamma(d/p))^2)^{-1/2}$.}. However, the resulting location hyper-parameter must be positive (if this condition can not be fulfilled a truncated normal prior should be used instead). The two set of hyper-parameter values shown in Table \ref{table:prior} for the generalised Gamma distribution result in positive and negatively skewed distributions, which for simplicity we call Generalised Gamma Right (GGR) and Generalised Gamma Left (GGL), respectively. Figure \ref{figure:prior} compares the three age prior distributions for the case of the $\beta$-Pictoris stellar association, where we use the $23\pm8$ Myr reported by \citet{2024MNRAS.tmp...30L}. As can be observed, the Generalised Gamma with positive skewness has a larger positive tail and a sharp cut at younger ages. 

As mentioned above, the three expansion rate components $\kappa_X$, $\kappa_Y$ and $\kappa_Z$ are drawn from the same statistical distribution, for which we use either a Normal or a StudentT, with location $\kappa_\mu$, scale $\kappa_\sigma$, and, in the case of the StudentT, degrees of freedom $\kappa_\nu$ (see Table \ref{table:prior}). The sampling can be done with two types of parametrization: central and non-central. In the central one, the parameters are directly drawn from the distribution while in the non-central one, first a latent random variable, $\vec{\eta}$, is sampled from the standardised distribution (i.e. location and scale set to 0 and 1, respectively), and then $\vec{\kappa}=\kappa_\mu +\vec{\eta}\cdot\kappa_\sigma$. This latter type of parametrization is sometimes useful to improve the performance of the NUTS algorithm. 

We include the Normal and StudentT distributions as prior for the $\vec{\kappa}$ components. While the Normal distribution helps to pull together information when the expansion is isotropic, the StudentT helps, with its long tails, in the non-isotropic cases by treating the most discrepant $\vec{\kappa}$ component as an outlier. This is typically the case of the $\kappa_Z$ component in NYSS older than 10 Myr due to the effects of the Galactic potential, which, in short time scales, acts primarily in this direction. For example, in $\beta$-Pictoris, both \cite{2014MNRAS.445.2169M} and \cite{2020A&A...642A.179M}, find expansion in the $\kappa_X$ and $\kappa_Y$ components and contraction in $\kappa_Z$, whereas in TW Hydrae, \cite{2023AJ....165..269L} finds isotropic expansion, with ages of 9.5 Myr, 9.8 Myr and 9.6 Myr, in the X, Y, and Z, directions, respectively. 

As a prior for the degrees of freedom parameter in the StudentT distribution, we use a Gamma distribution, which is the recommended one by \citet{doi:10.1198/jbes.2009.07145}. Here, we set the hyper-parameters $\alpha=2$ and $\beta=1$ resulting in a mode at dof=1, at which the StudentT reduces to the Cauchy distribution. The long tails of this latter will help in the non-isotropic expansion cases.

We set the Exponential distribution as prior for the $\kappa_\sigma$ parameter because it imposes a higher probability density to smaller dispersions. This Exponential distribution helps to minimise the dispersion between the three $\vec{\kappa}$ components, which is the expected behaviour if the association were expanding isotropically. We notice that rather than imposing isotropic expansion \citep[like, for example,][]{2019A&A...628A.123Z}, the Exponential prior simply specify this isotropy as the most probable case.

We included three types of family distributions as prior for the marginal position and velocity dispersions: the Truncated Normal, Exponential, and Gamma distributions. This latter is the default one in \textit{Kalkayotl}'s linear velocity model. We notice that these distributions are used as marginal prior for each of the three entries of the position and velocity dispersion vectors $\vec{\sigma}_{XYZ}$ and $\vec{\sigma}_{UVW}$. As hyper-parameter values for these prior distributions, we tested two options: a generic one and a informed one. The generic one, which is shown in Table \ref{table:prior}, uses the typical values for stellar associations and star-forming regions, which are 20 pc in position dispersion and 0.5 $\rm{km\,s^{-1}}$ in velocity dispersion \citep[see, for example, Table 9 of][]{2018ApJ...856...23G}. On the other hand, the informed option uses the a priori values of the system's position and velocity dispersions. Throughout the following sections we will use as informed phase-space dispersion prior the values shown in the last two rows of Table \ref{table:grid_parameters} (except in Sect. \ref{sensitivity:dispersion}). In general, the users of our expansion method can take as location values for these phase-space dispersion prior, those inferred by previous studies or with \textit{Kalkayotl}'s Gaussian model with joint positions and velocities \citep[see Sect. 2 of][]{2025A&A...693A..12O}. 

\section{Sensitivity analysis}
\label{sensitivity}

Once our Bayesian hierarchical model was specified, we perform a series of sensitivity analyses to choose its optimal configuration (i.e. the combination of prior distributions, parametrization, and hyper-parameters that renders the most performant age dating method). Once it is determined, we use it to evaluate its performance under some deviation conditions in which our method could be applied. To achieve these goals, we first define a series of metrics that we will use to objectively compare between different model configurations. Then, we apply our method with its various configurations to a set of synthetic datasets where the true age is known. We generated these synthetic datasets mimicking the properties of the benchmark $\beta$-Pictoris stellar association at its nominal age and distance (see Sect. \ref{data}). For simplicity reasons, we keep the fraction of observed radial velocity at 1.0. 

As metrics to evaluate our model we use the error, uncertainty, and credibility, all in percentage units. The error measures the relative deviation of the inferred posterior mean with respect to the true value, we notice that this error can be positive or negative depending if the inferred mean overestimates or underestimates the value. The uncertainty measures the relative standard deviation of the inferred posterior with respect to the true value. The credibility measures the fraction of synthetic simulations (for each particular case of the grid's parameters values) for which the true value is contained within the 95\% high-density interval (HDI) of the inferred posterior distribution. Finally, the optimal model configuration will be the one that minimises the error and the uncertainty and maximises the credibility, with priority on credibility, error, and uncertainty.

In the following sections, we first identify the optimal configurations of our model and then we use them to analyse their sensitivity to extreme conditions. First, we evaluate the model performance under shitting and scaling of the prior information. Then, we test the sensitivity of our model to varying phase-space dispersions of the stellar system. Finally, we test our model ability to recover the input true age under non-isotropic expansion.  These series of experiments will allow us to quantify the biases that our method will commit when applied to stellar systems deviating from the ideal conditions of unbiased prior age estimates,  well characterised velocity dispersions and isotropic expansion.

\subsection{Sensitivity to prior distribution and hyper-parameters}
\label{sensitivity:prior_hyper-parameters}

We apply the model configurations described in Sect. \ref{methods:prior} to the synthetic dataset of $\beta$-Pictoris. Then, we compared the results based on the previously described metrics and arrive at the following conclusions.

Concerning the position and velocity dispersions, we found that both hyper-parameter options (i.e. generic and informed) return similar metrics in all parameters, although the informed one probed to minimise the age error by $\sim$2\%, thus, we recommend its use. We notice that this option cannot be set as the default one in the \textit{Kalkayotl} code because it is system-specific, thus, the generic option remains the default. From the three tested family distributions for the position and velocity dispersions (i.e. Truncated Normal, Gamma, and Exponential), the ones that minimise the errors and uncertainties are the Gamma for positions and the Exponential for velocity. Therefore, we recommend their use and set them as default options. 

Regarding the scale hyper-parameter of the Exponential prior distribution of $\kappa_\sigma$, we tested values of $scale\!\in\![0.001,0.005,0.01]\, \rm{km\,s^{-1}\,pc^{-1}}$, which roughly correspond to age dispersions similar to 0.1, 0.5, and 1 Myr between the $\vec{\kappa}$ components. The three cases result in 100\% credibilities, low errors (<3\%), and uncertainties of 11\%, 13\%, and 15\% for the lowest to largest dispersions, respectively. Therefore, we recommend the use of the smallest dispersion and set it as the the default option. 

About the Normal and StudentT distributions that we use to sample the $\vec{\kappa}$ components, we observe that both result in 100\% credibilities, low errors (<2\%) and small uncertainties (<16\%). The only difference we notice between the results of these two distributions is that the Normal one produces lower errors (1\% vs. 2\%) in systems with 25 sources while in more populated systems the error resulting from the StudentT case diminishes. Therefore, we recommend the use of the Normal default configuration when there is evidence of isotropic expansion, which can be gathered by running \textit{Kalkayotl} with the linear velocity model without the expansion age module. In Sect. \ref{sensitivity:non-isotropic}, we will further evaluate the performance of these two distribution on stellar systems with non-isotropic expansion.  

With respect to the parametrization of $\vec{\kappa}$, we tested central and non-central cases, with them having errors <3\% and $\sim$7\%, respectively. The uncertainties of the central parametrization are 1\% smaller than the non-central one.
Therefore, we recommend the use of the central parametrization and we set it as the default one within \textit{Kalkayotl}.

\begin{figure*}[ht!]
\centering
\includegraphics[width=0.8\textwidth,page=1]{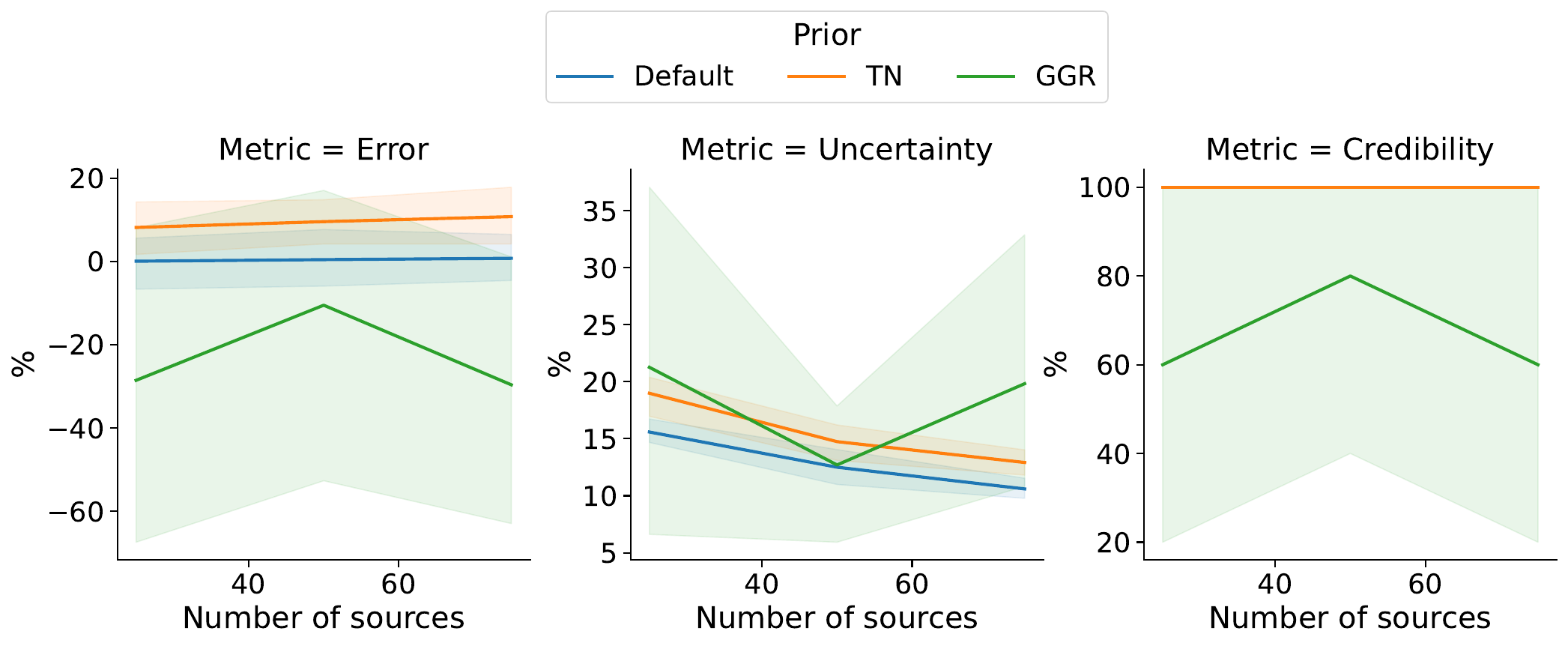}
\caption{Metrics for the age prior distributions for the generalised Gamma left (Default), truncated normal (TN), and generalised Gamma right (GGR). The lines and shaded regions depict the mean and standard deviation of the five synthetic simulations corresponding to each case.}
\label{figure:sensitivity:age_prior}
\end{figure*}

Figure \ref{figure:sensitivity:age_prior} shows the metrics of the age prior experiments. Concerning the tested prior distributions for the age parameter (i.e, TN, GGR and GGL), the GGR had convergence issues that resulted in large errors ($\sim$-30\%) and low credibilities ($\sim$80\%) most probably due to the sharp cut that it applies to younger ages (see, for example, Fig. \ref{figure:prior}). On the contrary, the TN and GGL resulted in 100\% credibilities, with the GGL showing the the smaller errors ($\lesssim$3\%) and uncertainties (<16\%), compared to those of the TN, which are 10\% in error and <19\% in uncertainty. Therefore, we conclude that the GGL distribution globally outperforms the TN. For this reason, we recommend its use and set it as the default option.

In the following, we will refer to the previous optimal configuration of hyper-parameters and prior distributions as the default age model, and we will denote it as $\rm{Default}(\cdot)$. Whenever we use a different configuration, we will explicitly state its difference from the default one. For example, when using a configuration in which the StudentT distribution is used to sample the components of the expansion rate instead of the Normal one, or one with non-central parametrization for the $\vec{\kappa}$ components, we will refer to them as $\rm{Default}(\vec{\kappa}\!\sim\!\rm{StudentT})$ or $\rm{Default}(\vec{\kappa}:\rm{non-central})$, respectively.

\subsection{Sensitivity to shifting and scaling of the a priori age information}
\label{sensitivity:shifting_and_scaling}

\begin{figure}[ht!]
\centering
\includegraphics[width=\columnwidth,page=6]{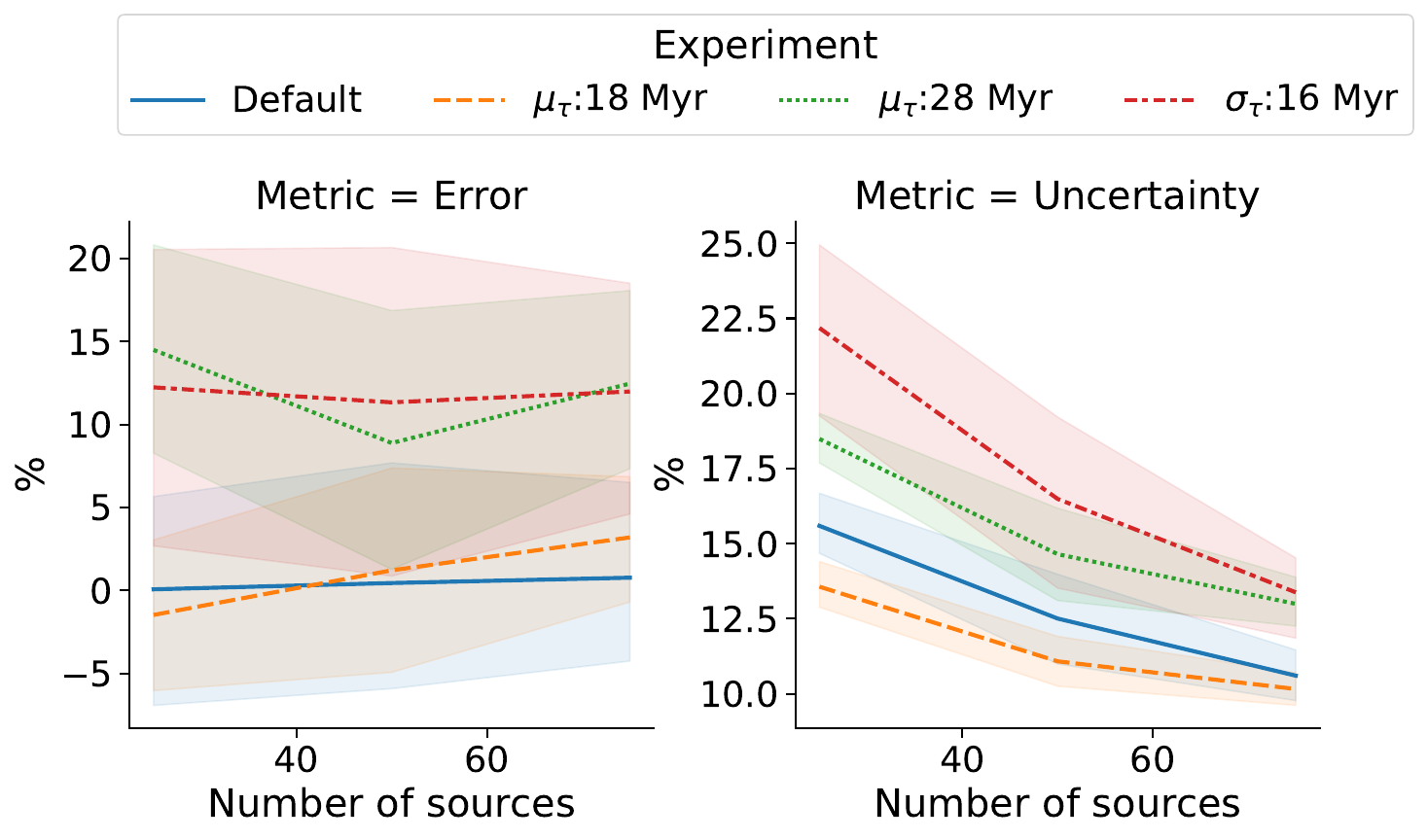}
\caption{Metrics for the shifting and scaling experiments. For comparison purposes, the default configuration without shifting and scaling is also included.  The lines and shaded regions depict the mean and standard deviation of the five synthetic simulations corresponding to each case. To save space, the credibility, which is always 100\%, is not shown.}
\label{figure:sensitivity:shift_and_scale}
\end{figure}

We also test the sensitivity our prior distributions to shifting and scaling of the prior age information, which represent the most general scenarios in which the literature age estimates can be obtained. These shifting and scaling can be done by modifying the location ($\mu_\tau$) and scale ($\sigma_\tau$) hyper-parameters from their nominal values of 23 Myr and 8 Myr, respectively. In the first hyper-parameter experiment, we shift the nominal location by $\pm5$ Myr and keep the scale at its nominal value. In the second experiment, we keep the location at its nominal value and doubled the scale to a value of 16 Myr.

Figure \ref{figure:sensitivity:shift_and_scale} shows the metrics of the previous experiments as a function of the number of sources. In these shift experiments, the errors have a sign that correspond to that of the input prior shift, with error values between 10\% and 15\% for the +5 Myr case and between -3\% and 3\% for the -5 Myr case. The uncertainties, as expected, reduce with increasing number of sources although their values differ. The larger uncertainties corresponds to that of the double scaled experiment and the positive shift, while the smallest to that of the negative shift.

Comparing the results from the different experiments, we observe the following. First, despite the symmetry in the prior shifts, the resulting errors are not symmetric, with the positive shift resulting in a larger bias than the negative shift, which returns almost negligible errors. This behaviour results from the asymmetry introduced by transformation from expansion rate to ages (see Eq. \ref{equation:expansion_to_age}) in which older ages are more probable than younger ones, despite this pull of the prior, the resulting estimate remains 100\% credible. Second, in the scale experiment we observe that doubling the prior scale has an impact in the error similar to that of a positive shift although with slightly larger uncertainty. This behaviour has its origin in the previously explained asymmetry of the transformation, with older ages being more probable than younger ones. Despite the 12\% error, the estimate remains 100\% credible in all random realisations.

From the previous experiments we conclude that whenever the a priori information on the system's age is scarce, shifting its mode towards younger ages should be preferred than to older ones. Nonetheless, as these experiments show, a prior shift by up to 20\% or an scale degradation by up to 70\% of the input age will result in credible age estimates. Despite these comforting results, we encourage the users of the age model to, whenever possible, infer ages with varying prior information and keep in mind that positive errors of 10-15\% could be expected.

\subsection{Sensitivity to phase-space dispersion}
\label{sensitivity:dispersion}

\begin{figure}[ht!]
\centering
\includegraphics[width=\columnwidth,page=8]{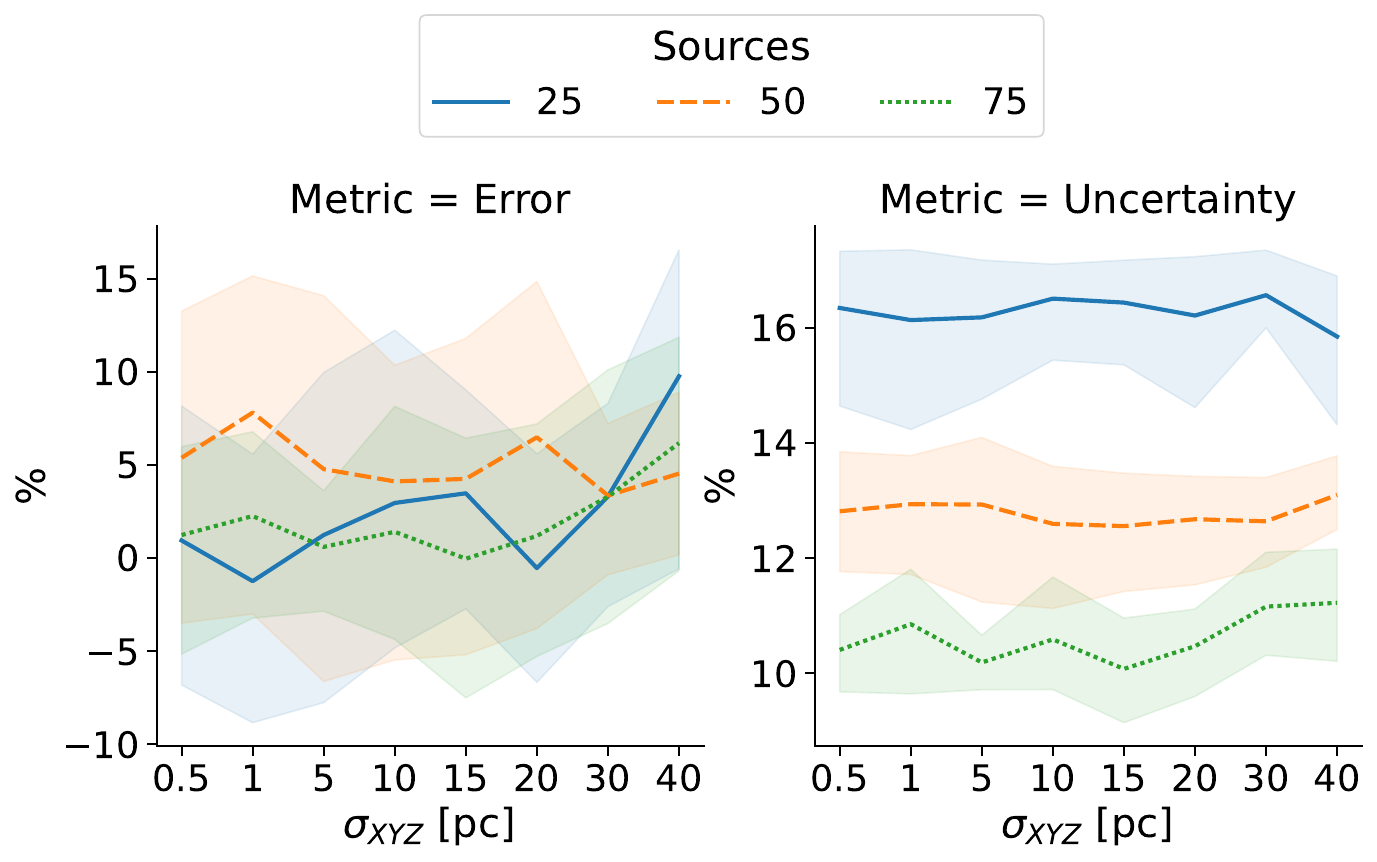}
\includegraphics[width=\columnwidth,page=9]{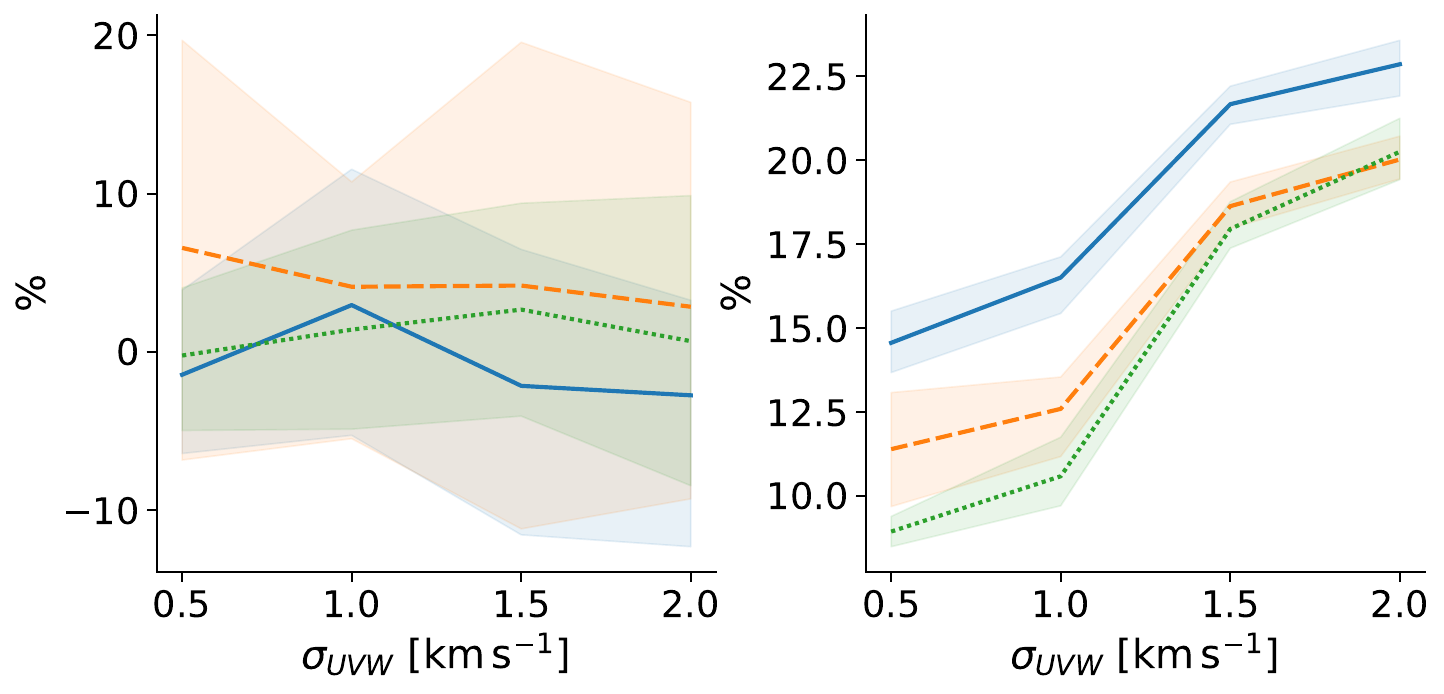}
\caption{Metrics for the phase-space dispersion experiments as a function of the position (top panel) and velocity (bottom panel) dispersion. The lines and shaded regions depict the mean and standard deviation of the five synthetic simulations corresponding to each case of the number of sources. To save space, the credibility, which is always 100\%, is not shown.}
\label{figure:sensitivity:dispersion}
\end{figure}

The NYSS can be found in a variety of phase-space configurations differing from those of the $\beta$-Pictoris and Taurus benchmarks. For this reason, we test the sensitivity of our improved Bayesian age estimator to the phase-space dispersion of stellar systems. To this end, we generated synthetic datasets based on the $\beta$-Pictoris template with nominal age and distance, but in which we vary the position dispersion, $\sigma_{XYZ}\in[0.5, 1, 5, 10, 15, 20, 30, 40]$ pc and the velocity dispersion, $\sigma_{UVW}\in[0.5, 1.0, 1.5, 2.0]\,\rm{km\,s^{-1}}$. Due to the varying phase-space dispersion values, we infer the model parameters using the default dispersion hyper-parameters for the Gamma and Exponential distributions for the position and velocity dispersions (see Table \ref{table:prior}). In addition and for simplicity reasons, we work with datasets having full radial velocity coverage. 

Figure \ref{figure:sensitivity:dispersion} shows the metrics of the inferred age in the various configurations described above. The figure shows the error and uncertainty metrics as functions of the position dispersion, $\sigma_{XYZ}$ (top panel) and velocity dispersion, $\sigma_{UVW}$ (bottom panel) for systems with 25, 50, and 75 sources. From this figure, we draw the following conclusions.

Concerning the position dispersion, $\sigma_{XYZ}$, we observe that the age errors are, on average, smaller than 7\% for systems with 50 sources, and smaller than 5\% for systems with 25 and 75 sources, with these values being uncorrelated with the magnitude of the dispersion up to values of 20 pc. As the dispersion increases beyond the scale length of our default uninformed prior, the errors increase from a few percent up to 5\% and 10\% for systems with 75 and 25 sources, respectively. The uncertainty shows no correlation with dispersion, and only a decrease with increasing number of sources, as expected. Finally, the age estimates are 100\% credible for all probed values of position dispersion. We notice that the age error in systems with 25 sources is smaller than that of systems with 50 sources. This effect is due to the effect of the prior, which is centred in the true value and dominates under these low-information content datasets.

Concerning the effects of the velocity dispersion, $\sigma_{UVW}$,  we observe that, similar to the case of the  position dispersion, the errors remain lower than 7\% for systems with 50 sources and <3\% for those with 75 and 25 sources. On the other hand, contrary to the trend observed in the position dispersion, the age uncertainty increases with increasing velocity dispersion. This effect is expected from the linear velocity model \citep[see Eq. \ref{equation:linear_model} and][]{2000A&A...356.1119L}, where the velocity of a star results from the addition of the velocity predicted by the linear velocity field plus a random value drawn from the Gaussian velocity dispersion (first and second terms in Eq. \ref{equation:linear_model}, respectively). Therefore, the constraining information provided by the datum of a single star has to be shared by these two terms. As the the velocity dispersion increases, the model needs an increasing information content to correctly estimate the velocity dispersion due to its Exponential prior (see Table \ref{table:prior}), which penalises large dispersions. As a result, there is less available information to constrain the linear velocity term, and thus, the system's age is less precise. For this reason, a constraining and fine tuned velocity dispersion prior will definitively help to increase the precision of the age estimate, as recommended in Sect. \ref{sensitivity:prior_hyper-parameters}.  

From the previous experiments, we draw the following conclusions. First, the age errors of our method are insensitive to the phase-space dispersion (both in positions and velocities) and only mildly to the number of sources. Second, the age uncertainty is sensitive to the velocity dispersion, varying from 10-15\% for systems with $\sigma_{UVW}\!\sim\!0.5\rm{km\,s^{-1}}$ to 19-23\% for systems with $\sigma_{UVW}\!\sim\!2.0\rm{km\,s^{-1}}$. Moreover, we expect that an informed prior in the velocity dispersion will help to further improve the age uncertainty.

\subsection{Sensitivity to non-isotropic expansion}
\label{sensitivity:non-isotropic}

\begin{figure}[ht!]
\centering
\includegraphics[width=\columnwidth,page=7]{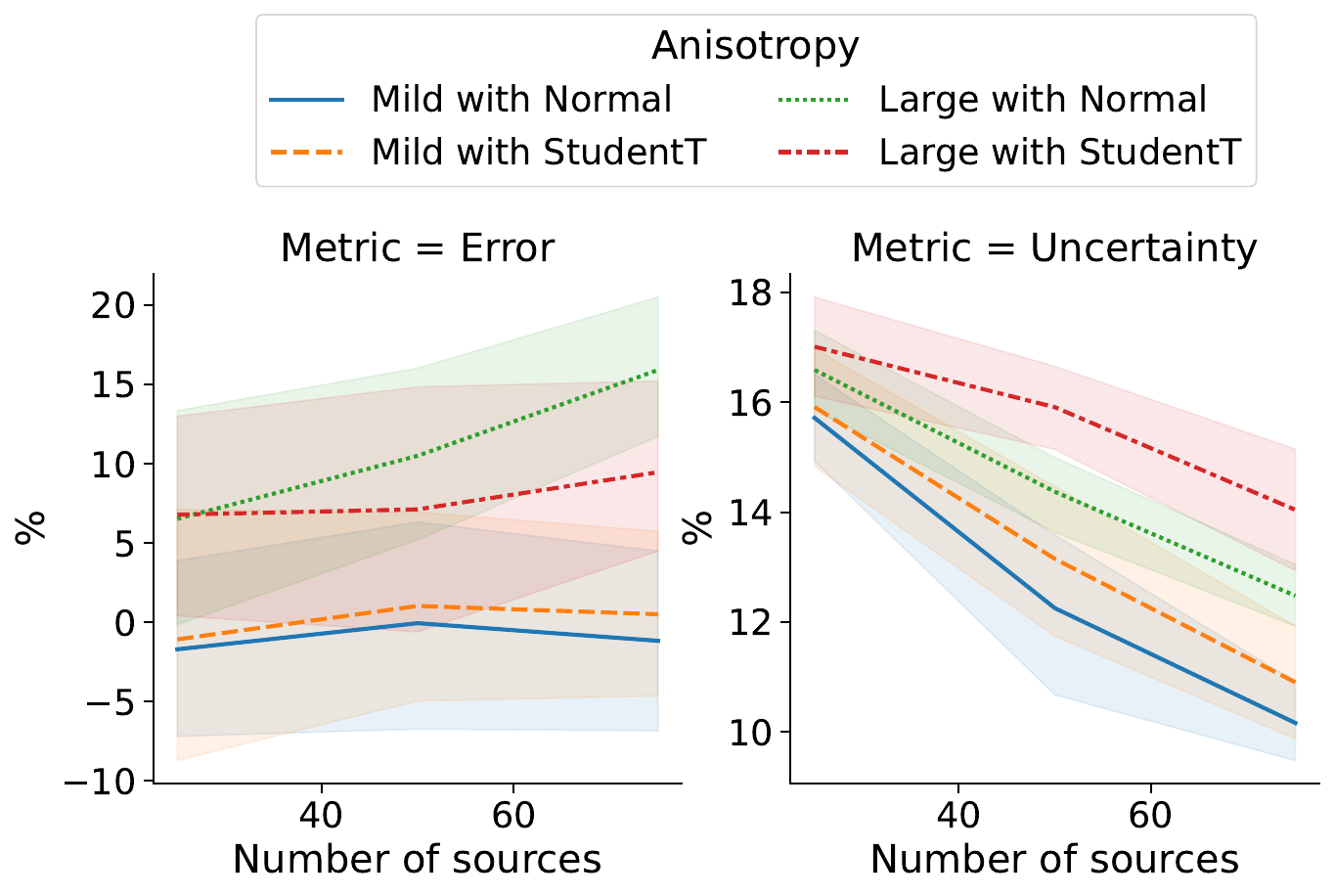}
\caption{Metrics for the non-isotropic expansion experiments. The lines and shaded regions depict the mean and standard deviation of the five synthetic simulations corresponding to each case. To save space, the credibility, which is always 100\%, is not shown.}
\label{figure:sensitivity:anisotropy}
\end{figure}

The expansion rate of stellar systems is not necessarily isotropic, and this anisotropy may originate from a variety of reasons, such as the geometry of the parent molecular cloud, the Galactic potential, or perturbations by nearby stellar systems or molecular clouds. Moreover, as explained in Sect. \ref{methods:prior}, the degree of anisotropy varies with the age of the system due to the accumulated effects of the non-symmetric Galactic potential. For all these reasons, we tested the sensitivity of our Bayesian dating method to non-isotropic expansion. To this end, we applied our method with its default and non-isotropic (i.e. $\rm{Default(\vec{\kappa}\sim StudentT)}$) configurations to synthetic stellar systems with non-isotropic expansion vectors. As before, we quantify the impact using the error, uncertainty, and credibility metrics. 

We generated stellar systems with non-isotropic expansion rates using the $\beta$-Pictors template at its nominal distance (see Sect. \ref{data:grid_parameters}). We created two experiments in which the degree of anisotropy was established varying the three $\vec{\kappa}$ components. In the first experiment, which we call mild anisotropy, the three expansion rate components disagree; we set the X, Y, and Z values so that they correspond (using Eq. \ref{equation:expansion_to_age}) to ages of 18 Myr, 23 Myr, and 28 Myr, respectively. In the second experiment, which we call large anisotropy, two of the entries agree while the other has negligible expansion. Thus, we set the X, Y, and Z values to correspond with ages of 23 Myr, 23 Myr, and 1 Gyr, respectively. This second experiment is intended to test our model in the more realistic scenario in which the Galactic potential has contracted the system along the Z direction, as in the case of $\beta$-Pictoris  \citep[see Sect. \ref{methods:prior} and ][]{2014MNRAS.445.2169M,2020A&A...642A.179M}.

Figure \ref{figure:sensitivity:anisotropy} shows the metrics obtained after inferring the ages of the two non-isotropic experiments with the default configuration (i.e. $\rm{Default(\mu_\tau=23,\sigma_\tau=8)}$) and the one optimised for non-isotropic expansion (i.e. $\rm{Default(\mu_\tau=23,\sigma_\tau=8,\vec{\kappa}\sim StudentT)}$, see Sect. \ref{methods:prior}). As can be observed from this figure, the default configuration returns negligible 2\% errors only in the mild anisotropy case where the value of the $\vec{\kappa}$ components disagree by an equivalent of $\pm5$ Myr from the location of the prior. However, in the large anisotropy experiment, we see that the default configuration results in errors varying from 5\% for the prior dominated case with 25 sources to 15\% in the data dominated case with 75 sources. Nonetheless, in spite of the huge difference in age (23 Myr vs 1 Gyr) between the components of the large anisotropy experiment, our method delivers age estimates that remain 100\% credible and with uncertainties between 13\% and 17\%. 

On the other hand, when the age is inferred with our model's non-isotropic configuration (i.e. $\rm{Default(\mu_\tau=23,\sigma_\tau=8,\vec{\kappa}\sim StudentT)}$), we observe the following improvements. In the mild anisotropy experiment, our model recovers the input age with smaller errors ($\sim$1\%) and slightly larger uncertainties than those recovered with the default configuration. In the large anisotropy experiment, we see a considerable reduction in the age errors with now reaching only 7\% in the data dominated case with 75 sources instead of the 15\% with the default configuration. Moreover, the error trend with increasing number of sources is also reduced resulting in an almost flat behaviour, indicating the stability and robustness of the non-isotropic configuration. The uncertainty in this case is larger by 2\% than in the default configuration, but we consider this a small price to pay considering the reduction in the errors and the large discrepancy in ages from the $\vec{\kappa}$ components. Finally, the non-isotropic configuration returns in both experiments a 100\% credibility.

From the previous experiments we can conclude that although both configurations return 100\% credible age estimates, the errors from the non-isotropic configuration are considerably reduced thanks to the robust statistical treatment given by the StudentT distribution. Finally, we recommend users of the age model to infer ages with the non-isotropic configuration whenever they have evidence of anisotropies in the expansion rate components. Nonetheless, biases of up to 10\% should be expected in the mean age determinations. 

\section{Validation}
\label{validation}
In this section, we present the validation of our new Bayesian expansion rate method (Sect. \ref{methods}) with its default configuration of prior distributions and hyper-parameters values (see Sect. \ref{sensitivity}; with the informed hyper-parameters for phase-space dispersion using values of Table \ref{table:grid_parameters}) evaluated on the grid of synthetic systems presented in Sect. \ref{data}. First, we present the validation on synthetic stellar associations, and then, we finish with the synthetic star-forming regions.

\subsection{Performance on stellar associations}
\label{validation:associtions} 

\begin{figure}[ht!]
\centering
\includegraphics[width=\columnwidth,page=1]{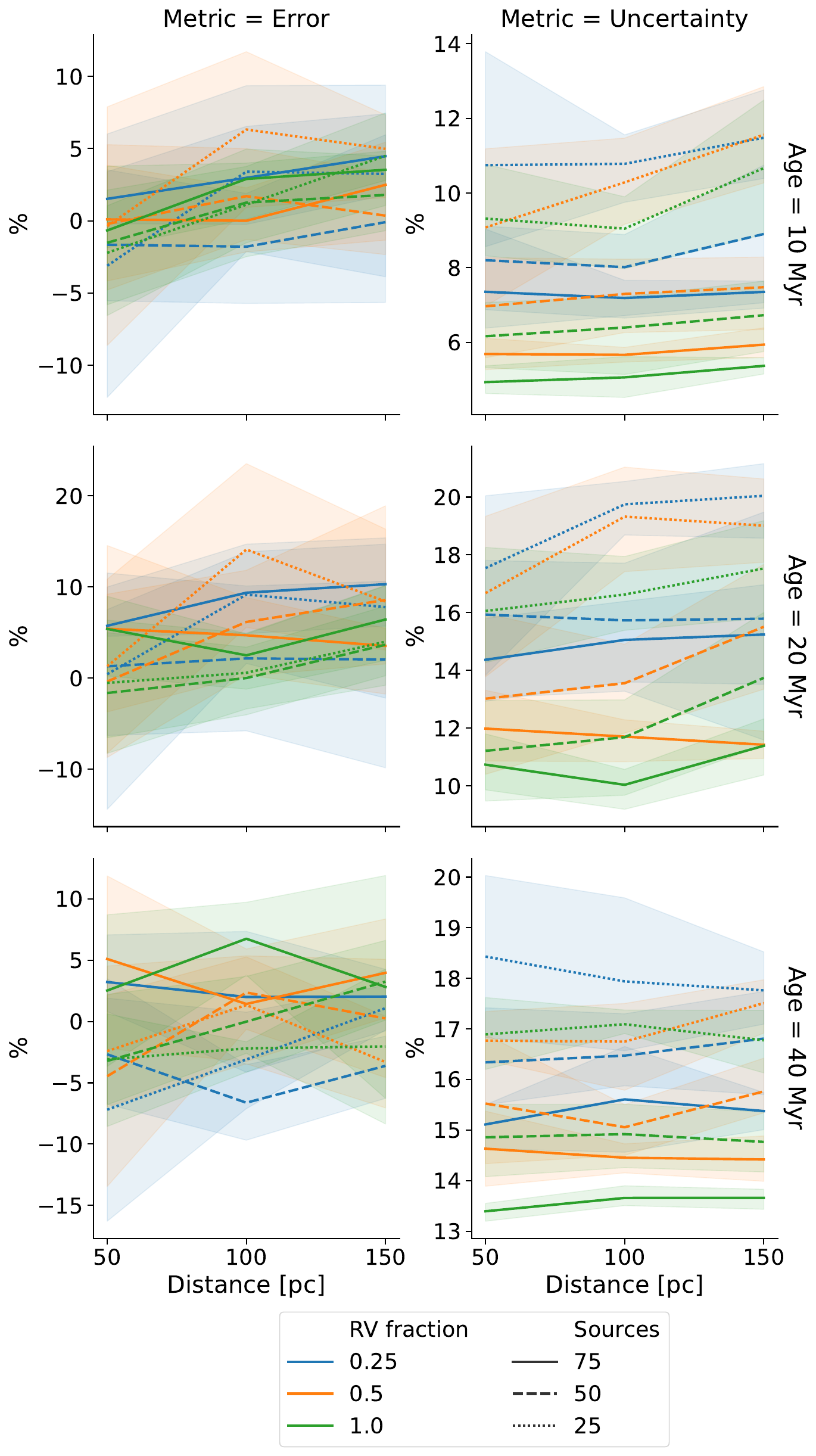}
\caption{Age's metrics of the synthetic associations as a function of distance, grouped by age (rows) and colour coded by number of sources and with line style showing the fraction of observed radial velocities. The lines and shaded regions depict the mean and standard deviation of the five synthetic simulations corresponding to each case. To save space, the credibility, which is always 100\%, is not shown.}
\label{figure:validation:associations}
\end{figure}

Figure \ref{figure:validation:associations} shows the metrics (as columns) of the Bayesian age estimator as a function of distance colour coded with the number of system's sources and grouped in three ages (as rows). The line style shows the fraction of observed radial velocities. At first glance, the errors are for all cases contained within $\pm10$\%, the uncertainties are below 20\%, and the credibility is always 100\%. The overall good value of these metrics over the entire applicability domain show the excellent properties of our improved expansion rate dating method.

Concerning errors, we observe that they increase with distance, particularly for datasets with low information content (i.e. those with 25 sources and 25\% of observed radial velocities). At the age of 10 Myr, there is no clear difference in the errors incurred in datasets with systems 25 to 75 sources although the errors slightly increase with distance. However, as age increases, there is a wider spread in age error that tends to be on the positive size with ages overestimated by 5-10\%.

Concerning the uncertainty, we observe an overall pattern in which its value increases with increasing distance, and decreases with increasing number of sources and fraction of observed radial velocities. Moreover, in the low information content datasets, this is those with 25 sources, we observe a larger dispersion in the metric's values compared to that of datasets containing 75 sources. The previous result highlights the importance of compiling membership lists with the largest number of system's members even when they lack radial velocities. Indeed, our results shows that doubling the number of sources even at the cost of halving the fraction of observed radial velocities provides better constraints to the system's age than those obtained in datasets with fully observed radial velocities but with half the number of sources.

Concerning credibility, our results show that the Bayesian age estimator provides 100\% credible ages in all tested conditions. This high quality metric results from the properly estimated uncertainties and low errors.

In summary, the series of experiments conducted on synthetic stellar associations show that the Bayesian age estimator  provides credible ages over its entire applicability domain. Moreover, its age estimates have errors <10\% and uncertainties <20\%.

\subsection{Performance on star-forming regions}
\label{validation:SF_regions}

\begin{figure}[ht!]
\centering
\includegraphics[width=\columnwidth,page=2]{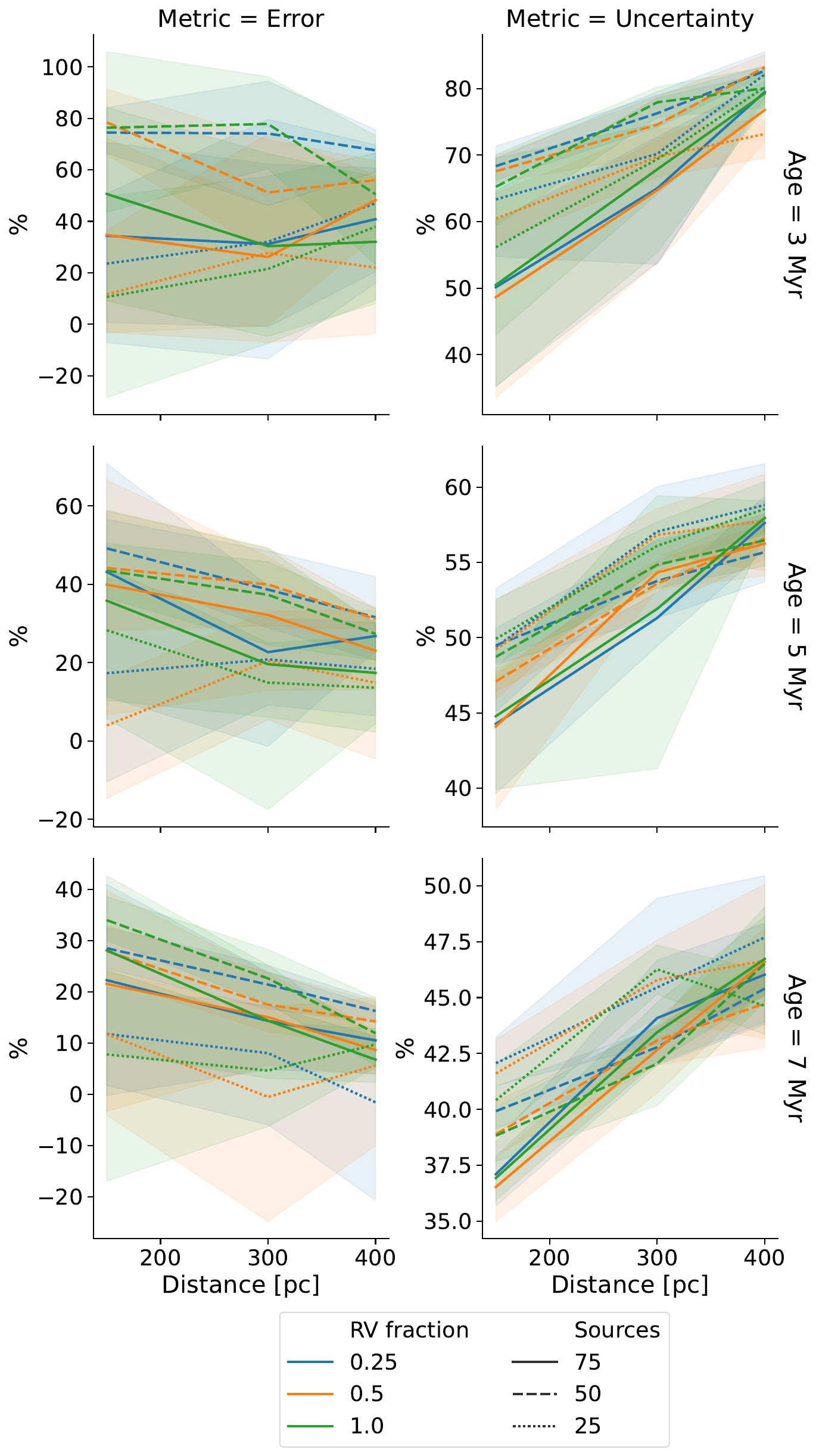}
\caption{Age's metrics of the synthetic star-forming regions. Captions as in Fig. \ref{figure:validation:associations}. To save space, the credibility, which is always 100\%, is not shown.}
\label{figure:validation:SF_regions}
\end{figure}

We applied our Bayesian expansion rate dating method to the series of synthetic star-forming regions presented in Sect. \ref{data}.  When applying our method to these datasets we notice that the default \textit{Kalkayotl}'s configuration for the NUTS sampling algorithm faced convergence issues. These issues resulted from both the lack of flexibility of the linear velocity model \citep[see Sect. 3.3 of][]{2025A&A...693A..12O} upon which the Bayesian age estimator is constructed and the phase-space compactness of the synthetic star-forming regions (see Sect. \ref{data:grid_parameters}). To overcome these issues, we increased number NUTS's sampling chains from two to four and discarded those with divergences, but always ensuring that at least two chains had converged. The metrics of the ages thus inferred are shown in Fig. \ref{figure:validation:SF_regions} from which we draw the following conclusions.

Concerning errors, we observe that these vary mostly with age and their dispersion gets reduced by increasing distance and lowering  number of sources. These last two effects are the consequence of the prior, which governs under low-information content datasets. The error's dispersion reaches its maximum at the closest distance and youngest age, where it varies from 0\% to a 80\%. As the age increases, the error dispersion gets reduced to a maximum of 50\% at 5 Myr and 35\% at 7 Myr.

Concerning uncertainties, we observe that these decrease with age and increase with distance, as expected. At 3 Myr, uncertainties vary between 50\% and 80\%, at 5 Myr between 45\% and 60\%, and at 7 Myr between 37\% and 47\%. 

Despite the large errors, the credibility of our age estimates remains at 100\%. Thus, in summary, our analysis on synthetic star-forming regions indicates that our Bayesian age estimator has a poorer performance when applied to star-forming regions compared to its results on stellar associations. Therefore, we advice caution when applying our method to the nearest and youngest star-forming regions, where large errors are expected. Moreover, we advice to always report the 95\% credible interval, which, on average, will cover the true age value of the stellar system. Nonetheless, future work will be needed to improve the flexibility of the linear velocity field model, and thus, reduce the errors of the Bayesian expansion rate dating method at the lower frontier of its applicability domain.

\section{Discussion}
\label{discussion}

Once our method was validated, we now discuss its advantages with respect to the frequentist approach and the possible caveats inherent to its Bayesian origin. Finally, we also discuss some possible future improvements. 

\subsection{Bayesian vs. frequentist age estimators}
\label{discussion:bayes_vs_frequentist}

In this section, we use the set of synthetic stellar associations (see Sect. \ref{data}, Table \ref{table:grid_parameters}, and Sect. \ref{validation:associtions}) to estimate ages with the classical and robust frequentist age estimators (see Sect. \ref{methods:frequentist}). In this cases, we increased the number of synthetic random simulations from five to ten. In these synthetic datasets, the data was first transformed to the physical space of positions and velocities, which implies discarding sources with missing radial velocities, and then the expansion rate components where independently estimated through a robust linear model regression in which parameter inference was done with the Random Sample Consensus algorithm \citep{10.1145/358669.358692} implemented in \textit{Scikit-learn} \citep{scikit-learn}. Then, ages were estimated through weighted means as explained Sect. \ref{methods:frequentist}. We then compare these frequentist ages to our Bayesian ones from Sect. \ref{validation:associtions} and discuss their differences based on the value of the error, uncertainty, and credibility metrics. 

\begin{figure*}[ht!]
\centering
\includegraphics[width=\textwidth,page=1]{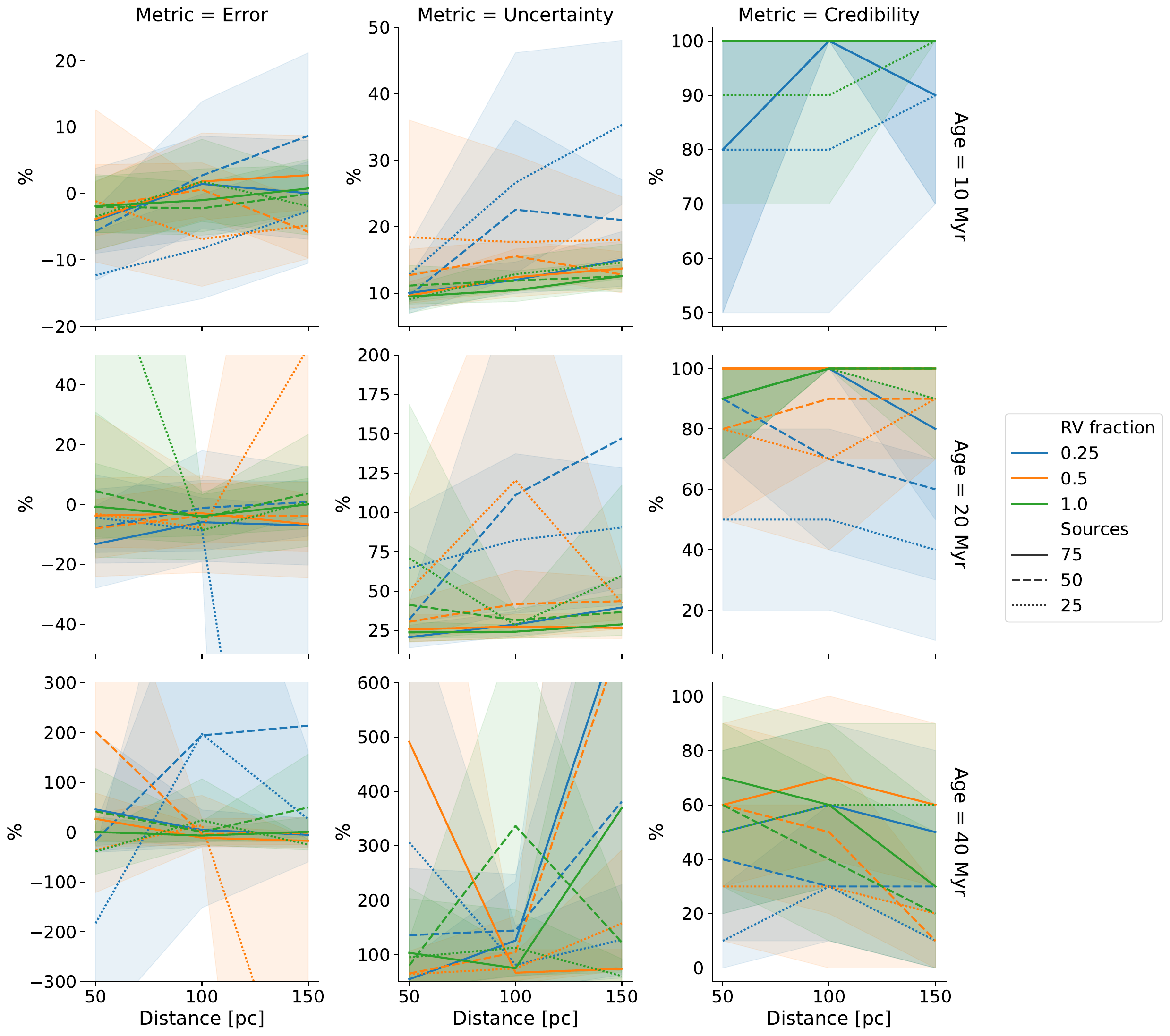}
\caption{Metrics of the classical frequentist age estimator applied to the datasets of synthetic associations. The lines and shaded regions depict the mean and standard deviation of the ten synthetic simulations corresponding to each case.}
\label{figure:classical}
\end{figure*}

\begin{figure*}[ht!]
\centering
\includegraphics[width=\textwidth,page=2]{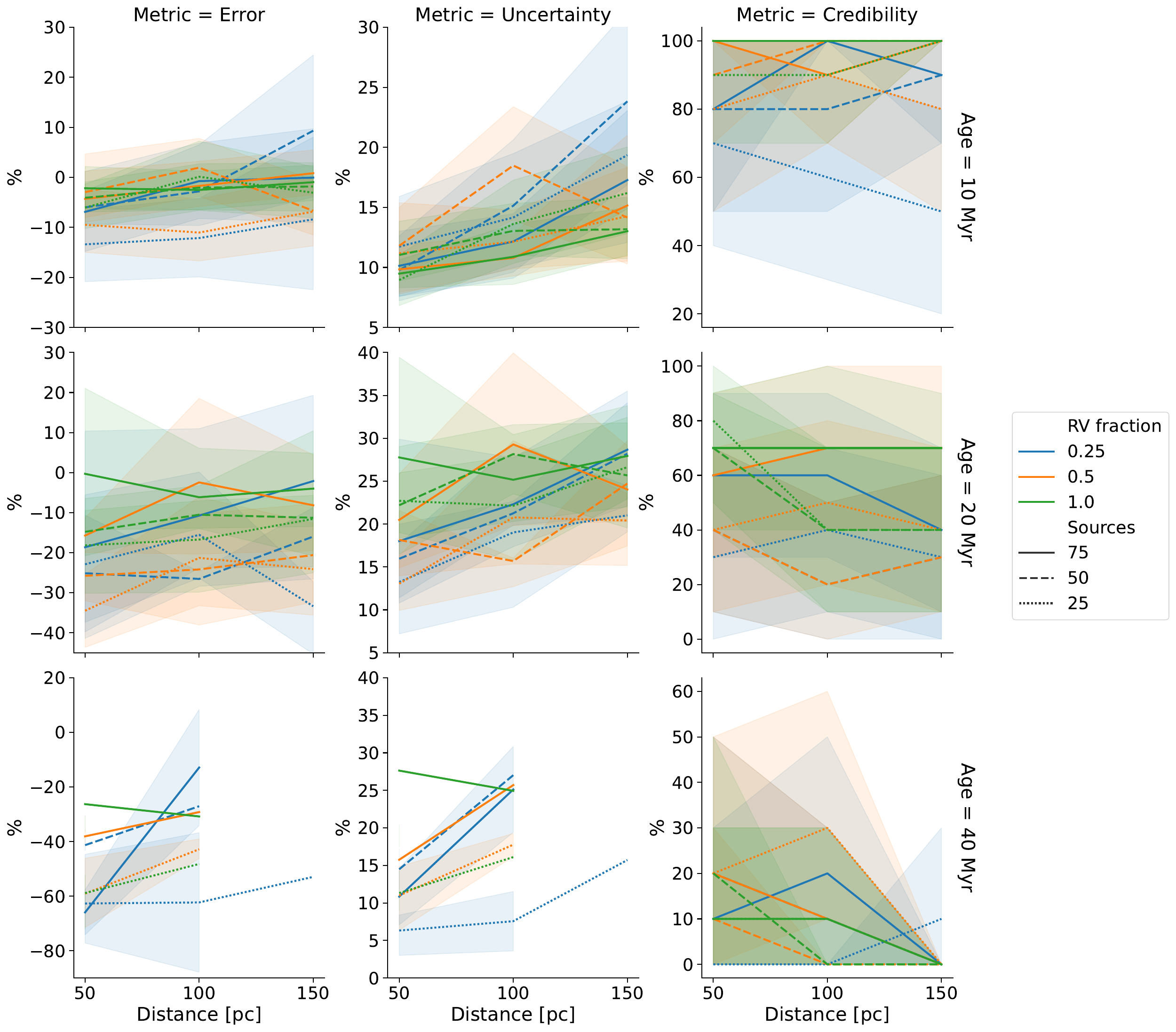}
\caption{Metrics of the robust frequentist age estimator applied to the datasets of synthetic associations. Captions as in Fig. \ref{figure:classical}.}
\label{figure:robust}
\end{figure*}

Figures \ref{figure:classical} and \ref{figure:robust} show the error, uncertainty and credibility metrics of the classical and robust frequentist age estimators, respectively. To simplify the comparison, the layout and captions in these figures correspond to those of the Bayesian estimator shown in Fig. \ref{figure:validation:associations}. From  the previous figures we draw the following conclusions.

Concerning the classical frequentist estimator (see Fig. \ref{figure:classical}), we observe that its errors vary with age, distance, number of sources, and fraction of observed radial velocities. We now describe its variation at each age value. 

At the youngest age of 10 Myr, the errors increase with decreasing fraction of observed radial velocities, varying from $<\pm2.5$\% for 1.0, $<\pm5$\% for 0.50, and $<\pm10$\% for 0.25. At this latter value, the ages are overestimated by 10\% at the largest distance of 150 pc and underestimated by -10\% at the closest of one of 50 pc. Nonetheless, in all cases, the errors decrease with increasing number of sources, as expected. The uncertainties are also reduced with increasing number of sources and increasing fraction of observed radial velocities, with values varying from 40\% at the farthest and less populous systems to 10\% at the closest and most populated ones. The credibility increase with increasing fraction of observed radial velocities, varying from 80-100\% at 0.25 to 90-100\% at 1.0. The low credibility in the closest 50 pc systems results from underestimated uncertainties.

At 20 Myr, we observe a degradation of the metrics' quality, particularly at the low-information content datasets with distance of 150 pc, 25 sources, and 0.25 fraction of observed radial velocities, where the error reaches values of -200\%, the uncertainty is larger than 75\%, and the credibility is lower than 50\%. As the fraction of observed radial velocity increases from 0.25, to 0.5 and 1.0, the errors get reduced to values between 0 and -15\%, -10\% and -5\%, respectively. Similarly, the uncertainties improve with decreasing distance and increasing fraction of observed radial velocities and number of sources. These uncertainties vary between 30\% and 120\% for systems with 25 sources, and get reduced to 20-40\% for systems with 75 sources. The credibility of systems with 25 sources is also smaller than in the 10 Myr case, with values of 40-50\%, 70\%-90\%, and 90\%-100\% for fractions of observed radial velocities of 0.25, 0.50, and 1.0, respectively. These credibilities increase with increasing number of sources, and reach 100\% in systems with 75 sources, except at the farthest distance.

At 40 Myr, the degradation of the metrics clearly indicates unreliable age estimates. The errors vary from $\pm200$\% to $\pm50$\% for fractions of observed radial velocities of 0.25 and 1.0, respectively, with only a mild improvement with increasing number of sources in the latter case. The uncertainties have erratic behaviours varying from 100\% to 500\% and no observable improvement with increasing number of sources nor fraction of observed radial velocities. As a consequence of these large errors and erratic uncertainties, the credibilities take values of 10-60\%, 20-60\%, and 20-60\% for fraction of observed radial velocities of 0.25, 0.50, and 1.0, respectively.

We notice that the quality of the metrics decreases with increasing age as a result of both the expansion velocity being inversely proportional to the system's age and the uncertainties of the observables remaining within the limits of the \textit{Gaia} properties. Thus, an older true age of the system implies a smaller expansion velocity and lower SNR of the expansion components. As a consequence, as age increases, the frequentist age estimate gets governed by errors.

Concerning the robust frequentist estimator (see Fig. \ref{figure:robust}), we observe that its behaviour is similar to the classical one although with smaller errors and improved uncertainties, which nonetheless result in reduced credibilities.

At the 10 Myr case, the errors mildly improve with the number of sources and decrease with increasing fraction of observed radial velocities. For the 0.25, 0.50 and 1.0 values of these latter, the errors take values of $\pm10$\%, $\pm7$\%, and $\pm5$\%, respectively. Similarly, the uncertainties improve mostly with increasing fraction of observed radial velocities, varying from 10-25\% at 0.25, 10-18\% at 0.50, and 9-16\% at 1.0. With respect to the credibility, it improves with increasing fraction of observed radial velocities and number of sources, taking values of 50-100\% , 80-100\%, and 90-100\% at fractions of observed radial velocities of 0.25, 0.5, and 1.0, respectively.

At 20 Myr, the three metrics are degraded with respect to the 10 Myr case, with now larger errors that underestimate the age in the majority of cases, and uncertainties and credibilities that mildly improve with increasing fraction of observed radial velocities and number of system's sources. Surprisingly, the ages are underestimated by 15\% to 35\% in systems with 25 sources and the errors are only reduced to the $\lesssim|10|$\% in systems with 75 sources and fully observed radial velocities. Furthermore, the uncertainties increase with increasing number of sources, which contradicts the expectations and the results of the classical estimator.  Observing the low credibility values in systems with 25 sources, we conclude that the uncertainty's unexpected behaviour results from underestimated values in poorly populated systems. Indeed, when the number sources increases, the uncertainties increase and the credibility increases as well, but only to reaches values between 60\% to 80\%. 

The 40 Myr case is by far the worst, with highly underestimated ages and uncertainties, resulting in credibilities that are never better than 30\%.  Moreover, in systems located at 150 pc, there is only one out of the ten simulations for systems with 25 sources and with a fraction of 0.25 of observed radial velocities in which the expansion rate component reaches the requested S/N>3. Similarly at 100 pc, only one out of the ten random simulation of systems with fully observed radial velocities fulfil the SNR criterion. Furthermore, if the SNR criterion would be increase it will further reduce the maximum distance at which the robust estimator returns a value.

Comparing the two frequentist age estimators we conclude that both are valid for systems with ages up to 10 Myr. From the two estimators, the classical one returns high credibility values despite the fraction of observed radial velocities whereas the robust one only does it for systems with large fraction of observed radial velocities. From this comparison, we also conclude that discarding one of the expansion rate components, as typically done with the Z component \citep[e.g.][]{2014MNRAS.445.2169M}, will effectively act as the robust estimator and result in an underestimated uncertainty, and as a consequence, a low credibility.

Comparing the metrics of the frequentist age estimators with those of the Bayesian one, we observe that the latter outperforms the former ones in the following aspects. First, its credibility is 100\% with 0 dispersion for all ages and cases, meaning that the true age value is always contained within the 95\% HDI whereas for the frequentist estimators this only occurs at the youngest ages with fully observed radial velocities. Second, contrary to the robust frequentist estimator and the classical one at 40 Myr, the Bayesian estimator has uncertainties that always behave as expected, with decreasing values for increasing number of sources. Moreover, the uncertainties of the Bayesian estimator show small changes $\sim$5\% with varying distance while those of the frequentist estimators vary greatly with age and distance, particularly at the old ages. Third, the Bayesian uncertainties are for all cases $\lesssim$20\%, which only occurs in the frequentist estimators for 10 Myr and fractions of observed radial velocities >0.5. Fourth, the errors of the Bayesian estimator are always smaller than those of the frequentist estimators, and only similar to these latter in the 10 Myr case.

The previous reasons allow us to conclude that, within the applicability domain of our method, our Bayesian age estimator outperforms the frequentist ones providing smaller errors, smaller uncertainties, and 100\% credibility. The only metric in which the frequentist estimators beat our Bayesian one is the computation time, where the former take minutes the latter take hours. 

\subsection{Caveats of the Bayesian approach}
\label{discussion:caveats}

Although the Bayesian age estimator outperforms its frequentist counterparts, it is not free of caveats and its underlying assumptions are not free of critics. The interested reader will find in Appendix \ref{appendix:assumptions} the list of our method's assumptions together with some basic criticism to them. Furthermore, as analysed and discussed in Sect. \ref{sensitivity}, the most important source of bias in our method is the a priori information embedded in the age prior, particularly when this information is biased towards older ages (see Sect. \ref{sensitivity:shifting_and_scaling}). However, in real case scenarios where the true system's age is not known, it is difficult to know, beforehand, if the a priori information is biased or not. Thus, in the following, we outline some recommendations to inspect the prior and to minimise its impact on the age's posterior distribution. 

\begin{figure}[ht!]
\centering
\includegraphics[width=\columnwidth,page=4]{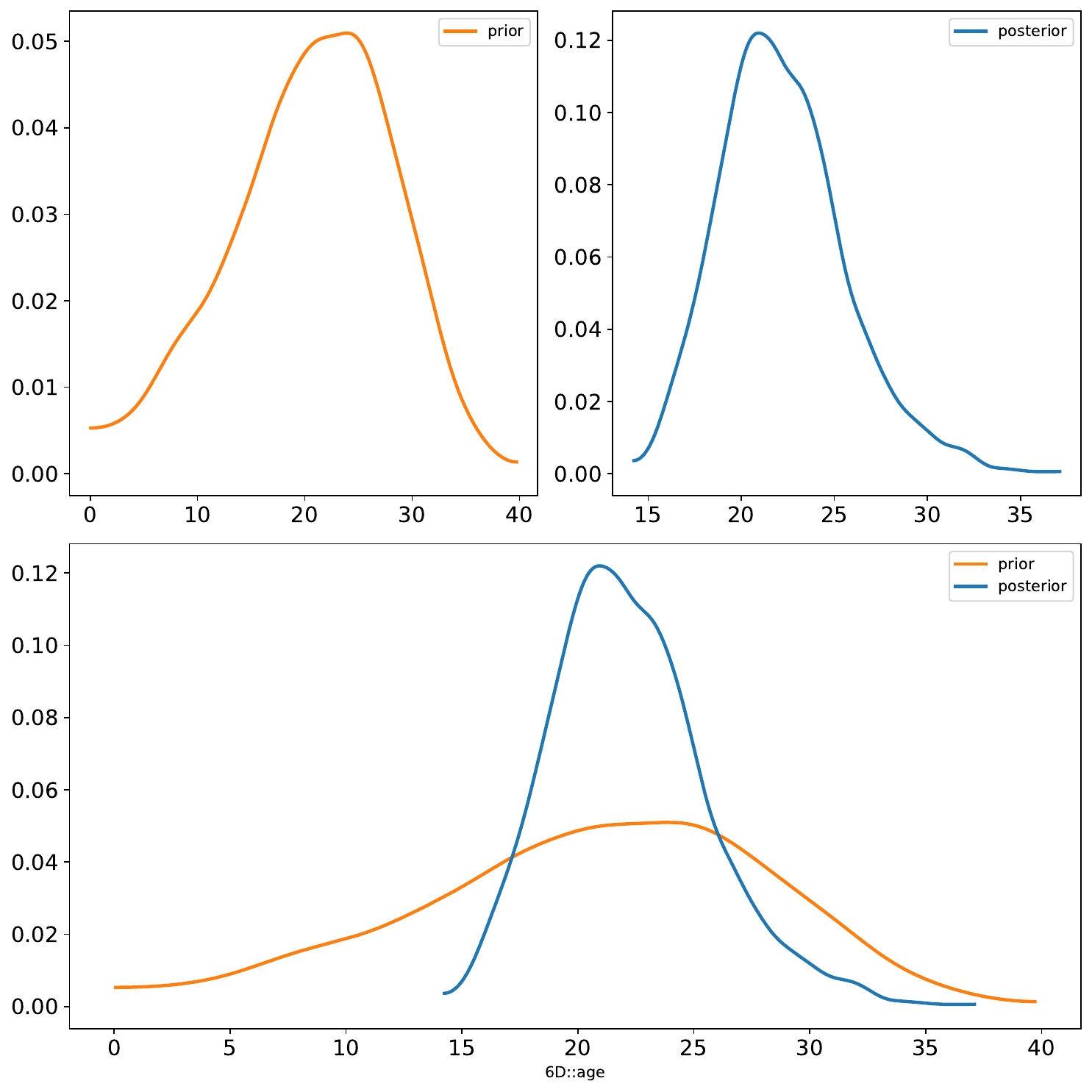}
\caption{Prior check plot of the age parameter (internally called 6D::age, in Myr) computed based of kernel density estimations from samples of the prior ($\mu_\tau$=23 Myr and $\sigma_\tau$=8 Myr) and posterior distributions for the $\beta$-Pictoris template with 25 sources and fully observed radial velocities. The top panels independently depict the two distributions while the bottom one plot them together.}
\label{figure:prior_check}
\end{figure}

\begin{figure}[ht!]
\centering
\includegraphics[width=\columnwidth]{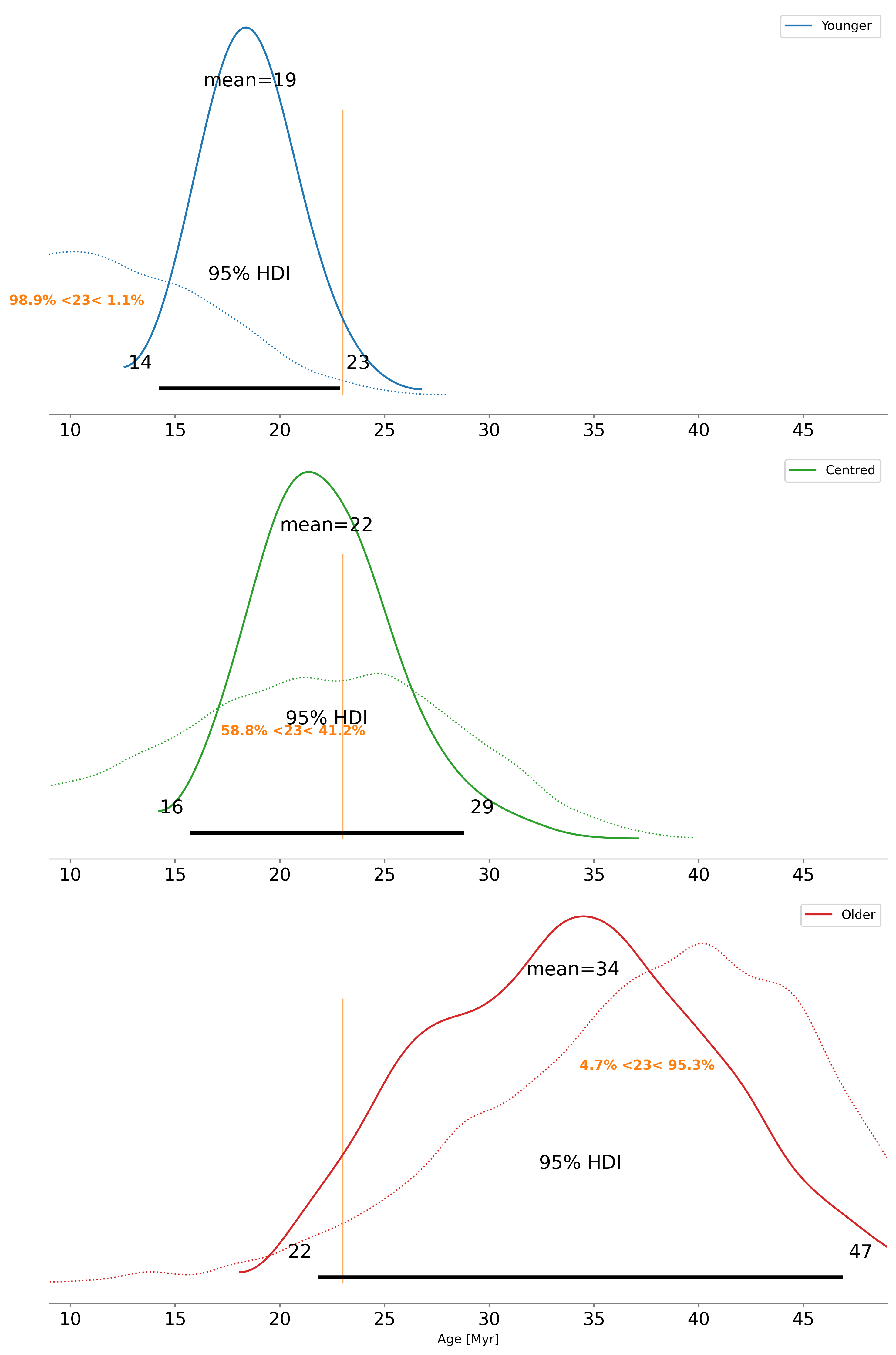}
\caption{Examples of the prior's influence on the posterior for the case of a realization of the $\beta-$Pictoris' template with 25 sources and fully observed radial velocities. The panels show kernel density estimates of the prior (dotted lines) and posterior (solid lines) distributions for the cases in which the prior is younger than (upper panel, $\mu_\tau=10$ Myr), centred at (mid panel, $\mu_\tau=23$ Myr), and older than (lower panel, $\mu_\tau=40$ Myr), the true system's age of 23 Myr (vertical orange lines). In all cases the prior dispersion is $\sigma_\tau=8$ Myr. The mean and 95\% HDI of the posterior are also shown in black text while orange one shows the percentages of the prior distribution that lay below and above the true age.}
\label{figure:prior_check_combined}
\end{figure}

First, it is fundamental to inspect the posterior distribution and compare it with the prior. Ideally, the prior should provide non-negligible probability in intervals where the posterior is not negligible. Moreover, the posterior should shrink with respect to the prior with the amount of this shrinking being proportional to the information content of the dataset. To help with this posterior inspection, \textit{Kalkayotl} provides, as part of its standard output, a file with figures in which the prior and posterior distributions of each model parameter are compared with each other. Figure \ref{figure:prior_check} shows an example of this type of figures in which the upper left and upper right panels show kernel density estimates obtained from samples of the prior and posterior distributions, respectively, and the lower panel combines the prior and posterior distributions into the same plot. Using these type of figures, the user can gauge  the impact that the prior has on the posterior. For example, in Fig. \ref{figure:prior_check_combined}, we compare the posterior distributions (solid lines) inferred for the $\beta$-Pictoris template (with 25 sources, fully observed radial velocities and a true age of 23 Myr), based on the following prior distributions (dotted lines): the younger prior (blue lines) with $\mu_\tau=10$ Myr and $\sigma_\tau=8$ Myr, the centred prior (green lines corresponding to Fig. \ref{figure:prior_check}) with $\mu_\tau=23$ Myr and $\sigma_\tau=8$ Myr, and the older prior (red lines) with  with $\mu_\tau=40$ Myr and $\sigma_\tau=8$ Myr. As can be observed, when the true age is barely contained by the prior the posterior barely include the true age, as in the case of the younger prior where only 1\% of it is above the true age and, and as a consequence, the majority of the posterior (i.e, the 95\% HDI goes from 14.2 to 22.9 Myr) does not include the true age. However, it suffices that only 5\% of the prior distribution contains the true age, for the 95\% HDI of the posterior to also do it, as in the case of the older prior. Thus, it is of paramount importance to criticise the prior under the light of the posterior and ensure that the majority of the latter is contained within the former. For example, in Fig. \ref{figure:prior_check} the prior provides non-negligible probability over the entire interval where the posterior is not negligible, thus ensuring that: i) the posterior freely moves on the parameter's space region constrained by the dataset, and ii) it shrinks with respect to the prior, as observed by the more compact effective domain (i.e. where the probability is not null).

Second, to avoid biasing the posterior with a restrictive prior, we recommend the use of weakly informative prior distributions \citep[e.g.][]{10.1214/08-AOAS191} in which the researcher deliberately provides the model with a prior distribution that is wider and less informative than the actual a priori information at hand. The weakly informative prior ensures not only better regularization properties but also reduces possible posterior biases. In the astrophysical context of age dating methods, this type of prior is particularly useful given that the a priori information typically comes from other dating technique, which are known to disagree \citep[e.g.][]{2022A&A...664A..70G,1999ApJ...522L..53B,1995AJ....109..298S} and underestimate uncertainties. Thus, our recommendation is to relax the a priori information by embedding a prior distribution with an increased scale.

Third, whenever there are some hints of possible posterior bias due to a restricting prior, we recommend to run the inference process with shifted or weakly informative (i.e. scaled) prior distributions until the recommendations of the first point are fulfilled. Doing this ensures that possible age biases remain minimal.

\subsection{Future methodological improvements}
\label{discussion:future_improvements}

From the modelling perspective, we foresee that the method would benefit from the following points. First, a phase-space model constructed on the system's reference frame (currently, it is only centred on it) and aligned with its natural axes. Second, a more complex phase-space model with second order velocity field, differential rotation, and differential expansion. Third, the inclusion of proper astrophysical phase-space distributions, as those presented in \citet{2018A&A...612A..70O}, rather than purely statistical ones. Our team is currently working on constructing phase-space models with second order velocity fields, on the natural reference frame of the system, and based on non-Gaussian multivariate statistical distributions and mixtures of them.   

From the astrophysical perspective, we believe that the nuclear and dynamical dating techniques will benefit from a joint and unified model that simultaneously infers the age of a system using several techniques and in which the systematics among these could be assessed on the basis of robust statistical comparisons. Our team is currently working on a Bayesian hierarchical model that simultaneously infers the age of a system through nuclear and dynamical ages and in which the zero-points offsets between techniques are inferred as free parameters of the model.
  
\section{Conclusions}
\label{conclusions}

The testing of current theories of star formation and stellar evolution demands accurate and precise ages of statistically significant samples of stellar systems. Due to their proximity and young ages, the NYSS have quality data making them excellent benchmarks to test these theories. Thus it is of paramount importance to estimate the ages of NYSS with the largest set of properly characterised dating techniques.

In this work, we improved the expansion rate dating technique with a Bayesian version of it that we make publicly available to the community. This Bayesian expansion rate method is rooted in the linear velocity field approximation whose parameters are inferred through a robust Bayesian multi level modelling framework. This Bayesian statistical framework allows us to: incorporate a priori information on the system's age and do a robust propagation of correlations and observational uncertainties from the data to the system's parameters, particularly the age. Using a large corpus of simulations, we prove that our Bayesian version of the expansion rate method produces age estimates with high quality metrics (i.e. high credibility and low error) that outperform those of the literature frequentist versions of the method. Nonetheless, our Bayesian age estimator could still be affected by random errors that depend on the system's properties and embedded age prior information.

\begin{acknowledgements}
We thank the anonymous referee for the provided comments and suggestions that greatly improved the quality of this manuscript.
JO acknowledge financial support from "Ayudas para contratos postdoctorales de investigación UNED 2021". "La publicación es parte del proyecto PID2022-142707NA-I00, financiado
por MCIN/AEI/10.13039/501100011033/FEDER, UE". 
A. Berihuete was also funded by TED2021-130216A-I00 (MCIN/AEI/10.13039/501100011033 and European Union NextGenerationEU/PRTR).
We express our gratitude to Anthony Brown, Jos de Bruijne, the Gaia Project Scientist Support Team, and the Gaia Data Processing and Analysis Consortium (DPAC) for providing the \textit{PyGaia} code.
We thank the \textit{PyMC} team for making publicly available this probabilistic programming language.

\end{acknowledgements}

\bibliographystyle{aa} 
\bibliography{bibliography} 

\begin{appendix}

\section{Assumptions and their criticism}
\label{appendix:assumptions}
\begin{assumption}
\label{assumption:expansion}
Young stellar associations are expanding. This assumption neglects possible contraction effects present in the very young stellar systems that may still be collapsing as well as zero-expansion rates in systems that are gravitationally bound.
\end{assumption}

Assumption \ref{assumption:expansion} can fail in very infant populations and in gravitationally bound systems. In very young systems, its members may still follow the gas dynamics, particularly the infall motion along filaments. Thus, expansion could be present in some spatial directions while contraction in others. Therefore, the age dating method would benefit from kinematic models constructed on the system's reference frame. In gravitationally bound systems, where expansion is expected to be null, the expansion rate method will simply fail providing an old age. In any case, whenever the total expansion signal is negligible other dating methods should be used instead.

\begin{assumption}
\label{assumption:expansion_at_birth}
The observed expansion rate of young stellar associations was imprinted at their birth. This assumption neglects possible time lags between the formation of the association's stars and the event that originated their expansion.
\end{assumption}

Assumption \ref{assumption:expansion_at_birth} could fail if the inferred expansion at the present day resulted from events occurring after the system's birth time. These events could include: i) supernova explosions, within the system or from a neighbour one, ii) encounters with massive objects, such as other stellar systems or giant molecular clouds, and iii) the gravitational potential removal due to gas expulsion. In all these cases, the expansion rate method would provide a younger age. Although this age would be biased, comparing it with age estimates from other dating techniques could provide valuable information to reconstruct the dynamical history of the system.

\begin{assumption}
\label{assumption:constant_expansion}
The observed expansion rate of young stellar associations has remained constant since their formation. This assumption neglects possible accelerations or decelerations of the association's expansion rate due to the Galactic gravitational potential or encounters with other stellar systems and molecular clouds.
\end{assumption}

Assumption \ref{assumption:constant_expansion} is violated by the presence of the Galactic gravitational potential. However, the accelerations imprinted by the Galactic potential start to distort the original shape of the system only after 40 Myr  \citep{1952BAN....11..414B}. Thus, the expansion rate method should remain valid until that age. Future work will be needed to review the validity of this assumption under a variety of realistic conditions.

\begin{assumption}
\label{assumption:single_population}
The phase-space distribution of a single population is described with a multivariate Gaussian distribution or a linear combination of them. In the latter case, we assume that the Gaussian distributions belong to the same population if and only if their mutual Mahalanobis distances are smaller than 2 (i.e. they are compatible at the 2$\sigma$ level).
\end{assumption}

Assumption \ref{assumption:single_population} could fail in systems with complex phase-space morphologies. However, the GMM distribution offers enough flexibility to accommodate non-Gaussian phase-space shapes, as shown by the substructures identified with this model \citep[e.g.][]{2023A&A...671A...1O,2023A&A...675A..28O}. The expansion rate method could benefit from a GMM in which each component has its independent linear velocity field.

\begin{assumption}
\label{assumption:unresolved_binaries}
The astrometry of unresolved binaries corresponds to that of their centre of mass and thus it can be used for the inference of the association parameters as if it were that of a single star. On the contrary, the observed spectroscopic radial velocity of unresolved binaries does not corresponds to that of their centre of mass. 
\end{assumption}

Assumption \ref{assumption:unresolved_binaries} is currently the best we can do to overcome the problems associated to unresolved binaries. In the future, additional information could be added to filter out these sources more efficiently in the data preprocessing steps.

\end{appendix}

\end{document}